\documentclass[acmsmall]{acmart}
\AtBeginDocument{%
  }

\usepackage{graphicx}
\usepackage{tikz}
\usepackage{mdframed}
\usepackage{tcolorbox}
\usepackage{enumitem}
\usepackage{subcaption}
\usepackage{xcolor}
\usepackage{algorithm}
\usepackage{algpseudocode}

\acmJournal{TKDD}

\begin{document}

\title{Integrating LLM-Derived Multi-Semantic Intent into Graph Model for Session-based Recommendation}

\author{Shuo Zhang}
\email{shuozhang@stu.ecnu.edu.cn}
\affiliation{%
  \institution{School of Data Science and Engineering, East China Normal University}
  \city{ShangHai}
  \country{China}
}

\author{Xiao Li}
\affiliation{%
  \institution{Samsung research China Beijing}
  \city{BeiJing}
  \country{China}}
\email{xiao013.li@samsung.com}

\author{Jiayi Wu}
\affiliation{%
  \institution{School of Data Science and Engineering, East China Normal University}
  \city{ShangHai}
  \country{China}
}
\email{wujwyi@126.com}

\author{Fan Yang}
\affiliation{%
 \institution{School of Electrical Engineering and Automation, Nantong University}
 \city{NanTong}
 \country{China}}
\email{fany@ntu.edu.cn}

\author{Xiang Li}
\affiliation{%
  \institution{School of Data Science and Engineering, East China Normal University}
  \city{ShangHai}
  \country{China}}
\email{xiangli@dase.ecnu.edu.cn}

\author{Ming Gao}
\authornote{Corresponding author.}
\affiliation{%
  \institution{School of Data Science and Engineering, East China Normal University}
  \city{ShangHai}
  \country{China}}
\email{mgao@dase.ecnu.edu.cn}

\renewcommand{\shortauthors}{Zhang et al.}

\begin{abstract}
    Session-based recommendation (SBR) is mainly based on anonymous user interaction sequences to recommend the items that the next user is most likely to click.  Currently, the most popular and high-performing SBR methods primarily leverage graph neural networks (GNNs), which model session sequences as graph-structured data to effectively capture user intent. However, most GNNs-based SBR methods primarily focus on modeling the ID sequence information of session sequences, while neglecting the rich semantic information embedded within them. This limitation significantly hampers model's ability to accurately infer users' true intention. To address above challenge, this paper proposes a novel SBR approach called Integrating LLM-Derived Multi-Semantic Intent into Graph Model for Session-based Recommendation \textbf{$\left( \text{LLM-DMsRec} \right)$}. The method utilizes a pre-trained GNN model to select the top-k items as candidate item sets and designs prompts along with a large language model (LLM) to infer multi-semantic intents from these candidate items. Specifically, we propose an alignment mechanism that effectively integrates the semantic intent inferred by the LLM with the structural intent captured by GNNs. Extensive experiments conducted on the Beauty and ML-1M datasets demonstrate that the proposed method can be seamlessly integrated into GNNs framework, significantly enhancing its recommendation performance. Implementation codes are available at  \url{https://github.com/nsswtt/LLM-DMsRec}.
\end{abstract}





\maketitle

\section{Introduction}
Recommendation systems are extensively utilized across online platforms such as music, gaming, and video streaming, providing an effective solution to address information overload. Traditional recommendation system methods primarily rely on users' historical preference data, focusing on modeling users' static preferences. However, the scarcity of resources and data makes it challenging to obtain users' historical preference profiles, thereby limiting the ability to accurately capture user intentions. In contrast, session-based recommendation (SBR) systems primarily rely on users' short-term interaction data, enabling the modeling of dynamic preferences and demonstrating strong applicability in addressing cold-start problems. 

The current mainstream SBR methods predominantly utilize graph neural networks (GNNs), such as SR-GNN\cite{wu2019session}, GCE-GNN\cite{wang2020global}, and DHCN\cite{xia2021self}. These methods model interaction sequences as graph-structured data and employ GNNs to capture the intricate preference patterns of users. Although GNN-based SBR methods have achieved superior recommendation performance, they primarily depend on the ID information of items in the sequence, neglecting the rich semantic information available.

\begin{figure}[t]
    \centering
    \includegraphics[height=5.48cm,width=7.5cm]{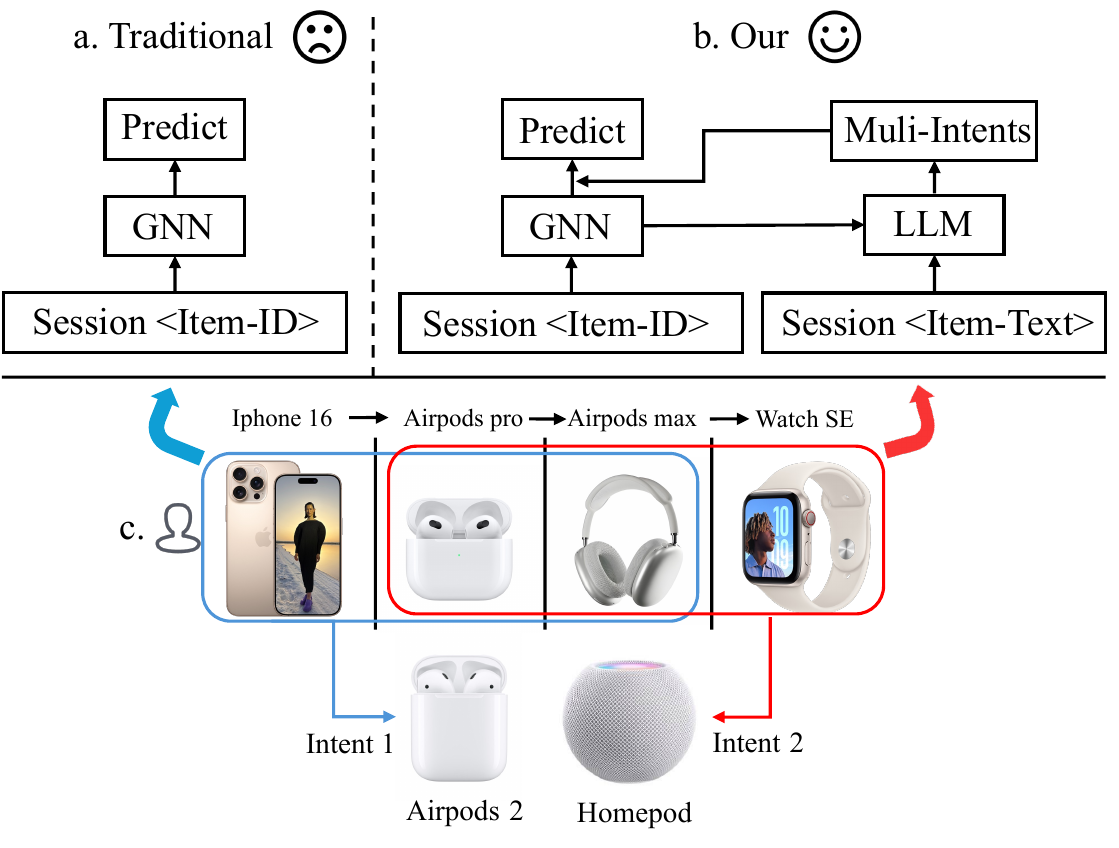}
    \caption{a. Traditional GNN-based SBR method; b. Our method; c. An example of a session sequence with multiple intents. Our method integrates the multi-semantic intent learned by LLMs into GNNs, significantly improving predictive performance in SBR.}\label{fig1}
    \Description{Three subfigures: a. Traditional GNN-based method; b. Our proposed method integrating multi-intent via LLMs; c. A session sequence example with diverse intents.}
\end{figure}

Recently, the emergence of large language models (LLMs)\cite{achiam2023gpt,bai2023qwen,touvron2023llama} has triggered major changes in research in both industry and academia. Its powerful text processing capabilities have found widespread applications across various domains of natural language processing (NLP)\cite{chowdhary2020natural}. The advanced language understanding and reasoning capabilities of LLMs have unveiled new avenues for research in SBR. Recent advancements have incorporated LLMs into SBR, exemplified by fine-tuning-based methods like LLMGR\cite{guo2024integrating} and reasoning-based approaches such as LLM4SBR\cite{qiao2024llm4sbr}. Among them, LLMGR leverages prompts designed with auxiliary and primary instructions to assist LLMs in comprehending graph-structured data and aligning textual information with corresponding nodes. LLM4SBR harnesses the reasoning capabilities of LLMs to capture users’ long-term and short-term intentions, integrating these insights with GNNs to enhance recommendation performance. Although the aforementioned LLM-based methods can utilize the semantic information of session sequences, they ignore the fact that a user's intentions within an interaction sequence are often dynamic and may encompass multiple facets. As shown in Figure \ref{fig1}-c, the user's click sequence of “\textit{Iphone 16 -> Airpods pro -> Airpods max -> Watch SE}” suggests that the first three interactive products indicate the user's potential intention to purchase headphones, such as “\textit{Airpods 2}”. In contrast, the last three interactive products suggest the user's intention may be to acquire a device connected to the phone, such as “\textit{Homepod}”. This prompted us to contemplate: \textit{In the context of user interaction session sequences, how can we leverage the powerful semantic understanding capabilities of LLMs to infer the multifaceted intentions of users?} In view of this thinking, we have identified the following key challenges:
\begin{itemize}[leftmargin=*]
    \item \textbf{How to accurately learn users’ multi-faceted intentions?} Within a session sequence, users may possess multi-faceted intentions, and it is of paramount importance to excavate the explicit and latent intentions of users from among these different intentions.  
    \item \textbf{How to alleviate the hallucination problem caused by LLM reasoning?} The user intent inferred by the LLM may be illusory or unrelated to any real intent represented in the item set.
    \item \textbf{How to align LLM tasks with GNN tasks?} LLMs leverage the text sequence of user interactions to learn semantic intent, while GNNs utilize the ID sequence to capture structural intent. Effectively and reasonably aligning these two tasks presents a significant challenge.
\end{itemize}

To address the aforementioned challenges, we propose LLM-DMsRec, a method that Integrating LLM-Derived Multi-Semantic Intent into Graph Model for Session-based Recommendation. Figures \ref{fig1}-a and \ref{fig1}-b illustrate a comparison between the SBR method based on the graph model and the framework proposed in this study. Our proposed method consists of three stages: (i) \textbf{Candidate Item Selection}, where a pre-trained GNN model is used to identify the top-K candidate item set, which serves as a knowledge base to alleviate the hallucination problem associated with user intent inference in LLM reasoning; (ii) \textbf{Intent Inference and Classification}, where we design prompts based on the candidate item set and session text sequence, and employ an LLM to infer the user's multi-semantic intents. Specifically, we classify the inferred intents into explicit and latent categories, and use a pre-trained BERT model to encode both types of intents; (iii) \textbf{Intent Alignment and Training}, where the encodings of both explicit and latent intents are integrated into the GNN model for joint training. Additionally, we introduce a KL divergence strategy to align the semantic and structural encodings of the two modalities.

Overall, the contributions of this work can be summarized as follows:
\begin{itemize}[leftmargin=*]
    \item We propose a multi-intention reasoning mechanism based on a LLM to extract the user's multi-semantic intentions from the candidate item set. To the best of our knowledge, our proposed method is the first to leverage LLMs to analyze users' multi-semantic intentions based on session semantic information.
    \item We developed an efficient intent-alignment mechanism and introduced KL divergence strategy to effectively align the semantic intent derived from LLM-based learning with the structural intent captured through GNN-based learning.
    \item Extensive experiments conducted on the Beauty and ML-1M datasets demonstrate that our proposed method, LLM-DMsRec, can be seamlessly integrated into the GNN model, resulting in superior recommendation performance.
\end{itemize}

\section{Related Work}

Next, we review the related work at three levels: traditional methods, deep learning approaches, and LLM-based techniques.

\subsection{Traditional-based approach}
Traditional SBR methods primarily include collaborative filtering\cite{wang2019collaborative,dias2013improving,park2011session,schafer2007collaborative}, Markov chain\cite{rendle2010factorizing,shani2005mdp}, and matrix decomposition\cite{liang2016factorization,he2016fast,anyosa2018incremental} approaches. 
Collaborative filtering-based methods primarily focus on the similarity between items within the session sequence to generate recommendations. Dias et al.\cite{dias2013improving} integrated temporal information into session-based collaborative filtering methods using both explicit and implicit strategies, achieving notable improvements in the accuracy of music recommendations. Building on the Markov chain framework, the user's behavior transition relationships are modeled to predict their next likely action. Shani et al.\cite{shani2005mdp} introduced a recommendation system model based on the Markov Decision Process (MDP), which was initialized using a predictive model and validated for its performance and effectiveness in real-world business scenarios.  Liang et al.\cite{liang2016factorization} proposed the CoFactor model, which learns user preferences by jointly decomposing the user-item interaction matrix and the item co-occurrence matrix. 
Although traditional SBR methods have demonstrated effectiveness, they fail to account for users' long-term intentions and the sequential relationships between item interactions.
        

\subsection{Deep Learning-based approach}
Deep learning-based SBR methods mainly include RNN-based\cite{hidasi2015session,tan2016improved,hidasi2018recurrent,moreira2020hybrid,dadoun2020many}, attention mechanism-based\cite{sun2019self,choi2024multi,luo2020collaborative,yuan2021dual}, and GNN-based\cite{wu2019session,wang2020global,xia2021self,ozbay2024gnn,peng2022gc} methods. Among these, RNN-based methods are primarily employed to capture the temporal dependencies between items in a sequence. Dadoun et al.\cite{dadoun2020many} introduced a many-to-one RNN model that predicts items a user is likely to click on by learning the probability distribution of action sequences performed within a session. The attention mechanism is primarily utilized to capture the key intentions of users. Choi et al.\cite{choi2024multi} introduced a self-attention mechanism to learn multi-level user intentions. GNNs are widely employed in session-based recommendation to effectively capture complex item dependencies within a session sequence. Wang et al. proposed the Global Context-Enhanced Graph Neural Network (GCE-GNN)\cite{wang2020global}, which integrates session-level and global-level item transition modeling to more effectively infer user preferences within the current session. Xia et al. proposed the Dual-Channel Hypergraph Convolutional Network (DHCN)\cite{xia2021self}, which leverages the synergy of hypergraph neural networks and self-supervised learning to substantially enhance recommendation accuracy. The deep learning-based approach, with its robust data modeling and feature extraction capabilities, more effectively captures user intentions, resulting in superior recommendation performance compared to traditional methods. However, most deep learning-based approaches focus on modeling ID sequences, neglecting the semantic information embedded in item attributes.

    

\subsection{LLM-based approach}
LLM-based methods can be divided into fine-tuning-based methods\cite{guo2024integrating,zheng2024adapting,wang2024llm,luo2024integrating} and non-fine-tuning-based methods\cite{qiao2024llm4sbr,wang2024re2llm,ren2024representation,sun2024large}. Fine-tuning LLMs on specific recommendation tasks has become a widely adopted approach. Guo et al. proposed the LLMGR framework\cite{qiao2024llm4sbr}, which integrates LLMs with GNNs to enable LLMs to better understand graph-structured data through the design of instruction-tuning task prompts. Wang et al.\cite{wang2024llm} proposed an innovative hint construction framework that transforms the relational information of graph data into natural language expressions, enabling LLMs to intuitively comprehend connectivity information within graph structures. While fine-tuning LLMs offers notable advantages for recommendation systems, it demands substantial computational resources and time investment. The LLM-based non-fine-tuning approach leverages pre-trained LLMs without additional fine-tuning. Qiao et al. proposed LLM4SBR\cite{qiao2024llm4sbr}, which reasons over session text data from multiple perspectives and fuses session representations across different viewpoints and modalities for recommendation.
Wang et al. proposed the Reflection-Enhanced Large Language Model (Re2LLM)\cite{wang2024re2llm}, which improves recommendation accuracy by incorporating a reflective exploration module for reasoning guidance and a reinforcement utilization module for reasoning refinement. Ren et al.\cite{ren2024representation} introduced a recommendation paradigm that integrates representation learning with LLMs, enabling the capture of complex semantic aspects of user behaviors and preferences. Although the LLM-based non-fine-tuning approach can process textual information in recommendation tasks, it lacks task-specific optimization, resulting in less accurate reasoning. Additionally, it heavily depends on effective prompt design, limiting its ability to fully capture user intentions.

To comprehensively mine the user's multi-intentions at the semantic level, we propose \textbf{LLM-DMsRec}. For the user interaction text sequence, we leverage an LLM and the candidate item set to infer the user's multi-semantic intentions. These intentions are further categorized into explicit and latent intentions, which are then aligned with the structural intentions learned by the GNN for the recommendation process.

\section{Preliminaries}
Next, we will introduce the preparatory work of this paper from two perspectives: problem definition and prompt design. 


\begin{tcolorbox}[colback=white, colframe=blue!75!black, title=Prompt 1: \textit{Instruction Prompt}]  
    \textit{You are tasked with inferring the user’s intents based on a sequence of items they have interacted with}.\\
        
    \textit{\textbf{Requirements}:}
        
        
        
    \begin{enumerate}[leftmargin=0.5cm]
    \item[1.] \textit{If multiple intents are inferred, list all relevant intents that reflect the user's current preferences and separate them with semicolons.}
    \item[2.] \textit{The inferred intents must be selected from the Candidate item set.}
    \item[3.] \textit{Note that the number of recommended intents should be appropriate. } 
    \end{enumerate} 
    \end{tcolorbox}  

    \begin{tcolorbox}[colback=white, colframe=red!75!black, title=Prompt 2: \textit{Input Prompt}]  
    \textit{\textbf{The order in which users click on items is as follows:}} 

    \begin{enumerate}[leftmargin=0.5cm]
    \item[1.] \textit{Zia Bamboo Exfoliant, 1.6 Ounce Bottles (Pack of 12)\_2664;}
        
    \item[2.] \textit{Remington S1051 Salon Quality, Professional Ceramic Hair  Straightener\_4151;}
        
    \item[3.] \textit{Sunburnt Therapeutic After Sun Relief, 6-Ounce Tube\_3889;}
        
    \item[4.] \textit{Vaseline Men Body and Face Lotion , 20.3 Ounce Bottle (Pack of 3)\_3714;}
        
    \item[5.]\textit{Cotz Pediatric Spf 40, 3.5 Ounce\_10164\\}
    \end{enumerate} 
        
    \textit{\textbf{Candidate item set:}}

    \begin{enumerate}[leftmargin=0.5cm]
    \item[1.] \textit{Olay Regenerist 14-Day Skin Intervention}
        
    \item[2.] \textit{Max Factor Panstik Foundation - 13 Nouveau Beige}
        
    \item[3.] \textit{MoroccanOil Hydrating Styling Cream, 10.2-Ounce Bottle}
        
    \textit{...}
    \item[50.] \textit{Jane Carter Solution Wrap and Roll, 8 Oz}
    \end{enumerate} 
    \end{tcolorbox}

    \subsection{Problem definition}
        SBR aims to predict the item a user is most likely to interact with next, based on the sequence of anonymous user interactions within a given session. Let $V=\left\{v_{1},v_{2},...,v_{n} \right\}$ denote the set of all items in the dataset, where $n$ represents the total number of items. Additionally, let $S=\left\{s_{1},s_{2},...,s_{m} \right\}$ represent the sequence of all user interaction sessions, where $m$ denotes the total number of sessions in the dataset. Here, $s_{t}=\left ( v_{1},v_{2},...,v_{l} \right )$ represents the session at time $t$, where $l$ denotes the number of items in the session $s_{t}$. Given a session $s_{t}=\left ( v_{1},v_{2},...,v_{l} \right )$ the goal of SBR is to predict the next item, $v_{l+1}$, that is most likely to be clicked. Formally, the recommendation task can be framed as an optimization problem, which is expressed as follows: 
        \begin{equation}
            \hat{y}=arg\underset{y\in \mathcal{Y}}{max}P\left ( y|s_{t};\Theta \right )=arg\underset{y\in \mathcal{Y}}{max}P\left (s_{t};y;\textbf{H};\Theta \right )
        \end{equation}
        Here, $\mathcal{Y}$ denotes the set of candidate items, $s_{t}$ represents the user's interaction sequence at time t, $\textbf{H}$ corresponds to the user's intention embedding matrix, and $\Theta$ represents the model parameters.

    \subsection{Prompt design}
        To fully leverage the language understanding and reasoning capabilities of LLMs while tailoring them to the specific requirements of SBR tasks, we propose a general interaction paradigm with LLMs. This paradigm consists of two key components: \textbf{Instruction prompts} and \textbf{Input prompts}. Instruction prompts guide LLMs to infer users' multi-semantic intentions, while input prompts transmit the text sequence and candidate item set derived from user interactions to the LLMs. For instance, given a user's chronological interaction sequence: $v_1$ : “\textit{Zia Bamboo Exfoliant, 1.6 Ounce Bottles (Pack of 12)}”, $v_2$ : “\textit{Remington S1051 Salon Quality, Professional Ceramic Hair Straightener}”, $v_3$ : “\textit{Sunburnt Therapeutic After Sun Relief, 6-Ounce Tube}”, $v_4$ : “\textit{Vaseline Men Body and Face Lotion, 20.3 Ounce Bottle (Pack of 3)}”, $v_5$ : “\textit{Cotz Pediatric SPF 40, 3.5 Ounce}”, the instruction prompt assists the LLM in deducing beauty products that users may be interested in by providing specific instructions and contextual guidance. Simultaneously, the input prompt facilitates the interaction by feeding the LLM with the user's text sequence and candidate item set.
        
        Prompt 1 and Prompt 2 serve as illustrative examples for the construction of the Instruction Prompt and Input Prompt on the Beauty dataset, respectively. Notably, their design remains consistent when applied to the ML-1M dataset.

\section{Methodology}
\begin{figure*}[t]
    \centering
    \includegraphics[height=8cm,width=10cm]{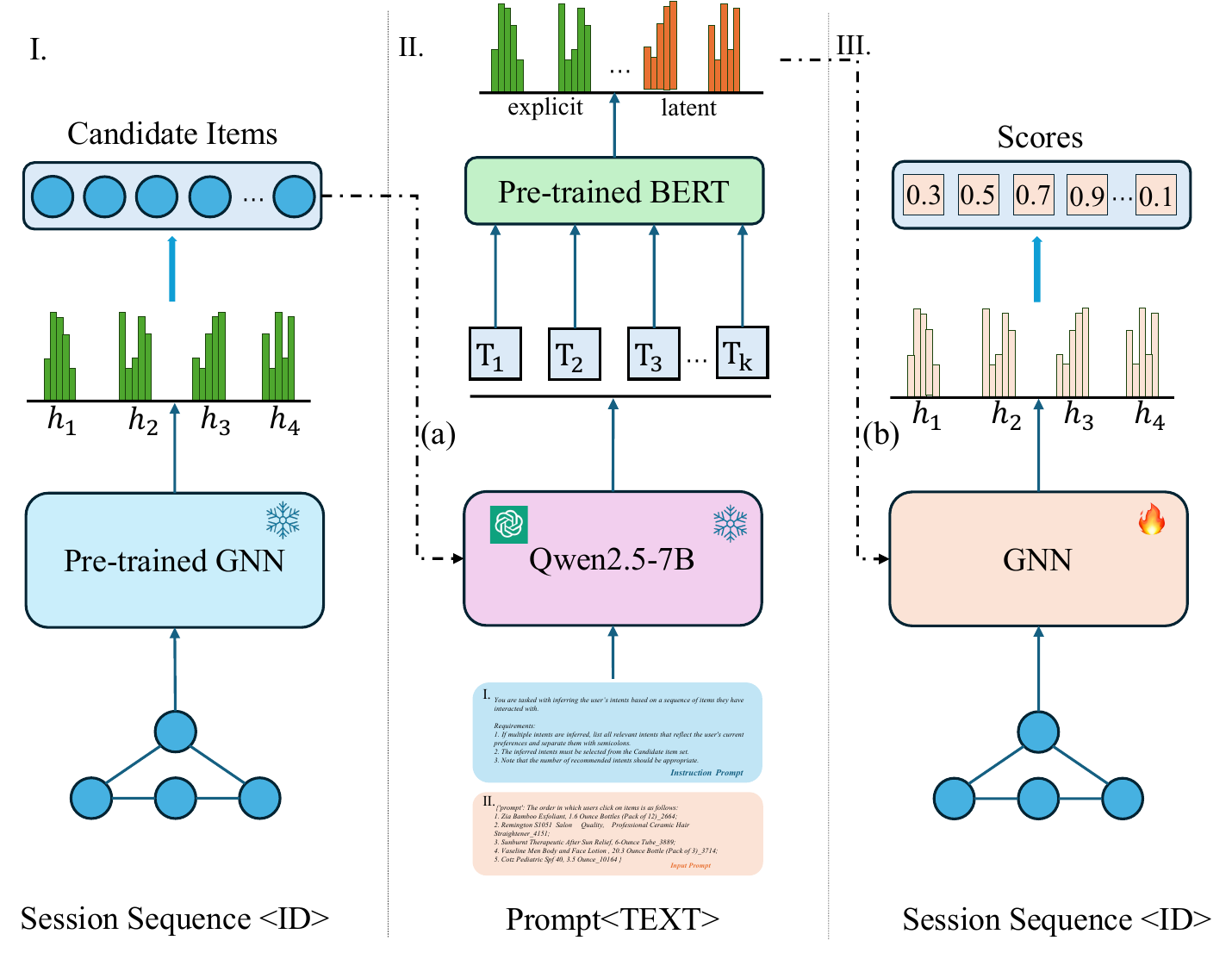}
    \caption{The overall architecture of the LLM-DMsRec model. Phase I employs a pre-trained GNN model to derive the candidate item set from the session ID sequence. Phase II utilizes the candidate item set and LLM to infer the user's multiple intentions, incorporating embeddings from a pre-trained BERT model. Phase III integrates the inferred semantic intentions into the GNN model and conducts joint training. (a) Represents mapping the candidate item set ID sequence to the text sequence and using it as the knowledge base for LLM reasoning. (b) Indicates the integration of explicit and latent intent features into the GNN model.}\label{fig2}
    \Description{}
\end{figure*}

    In this section, we present our proposed method, \textbf{LLM-DMsRec}, from three key perspectives: candidate item set acquisition, intent reasoning mechanism, and intent alignment and training. Following this, we detail the prediction and optimization processes of the method.  Figure \ref{fig2} shows the overall structure of LLM-DMsRec.

\subsection{Candidate Item Set}
The candidate item set serves as the knowledge base for the LLM to infer user intentions, thereby enabling the model to focus more effectively on areas aligned with user interests, while mitigating the hallucination problem during LLM reasoning. Specifically, given the interaction ID sequence \(s_t = (v_1, v_2, \ldots, v_l)\) of the user at time \(t\), a pre-trained GNN model is employed to select the top-\(K\) highest-ranked items as the candidate item set. This process can be formalized as follows:
\begin{equation}
    \{I_1, I_2, \ldots, I_k\} = \text{Pre-trained GNN}(s_t, \mathbf{W})
\end{equation}
In the context of the GNN model, \( \textbf{W} \) denote the pre-trained parameters. \( I_k \) represent the ID of the \( k \)-th ranked item. At this stage, \( I_k \) is still the item ID. To obtain the corresponding textual information for each item, we apply the Text-Mapping function, which maps the item IDs to their respective text titles. Specifically, we have:
\begin{equation}
    \{T_1, T_2, \dots, T_k\} = \text{Text-Mapping}(\{I_1, I_2, \dots, I_k\})
\end{equation}
where \( T_k \) denotes the textual title information of the \( k \)-th item.

\subsection{Intent Reasoning Mechanism}
\begin{figure}[t]
    \centering
    \includegraphics[height=5.48cm,width=7.5cm]
    {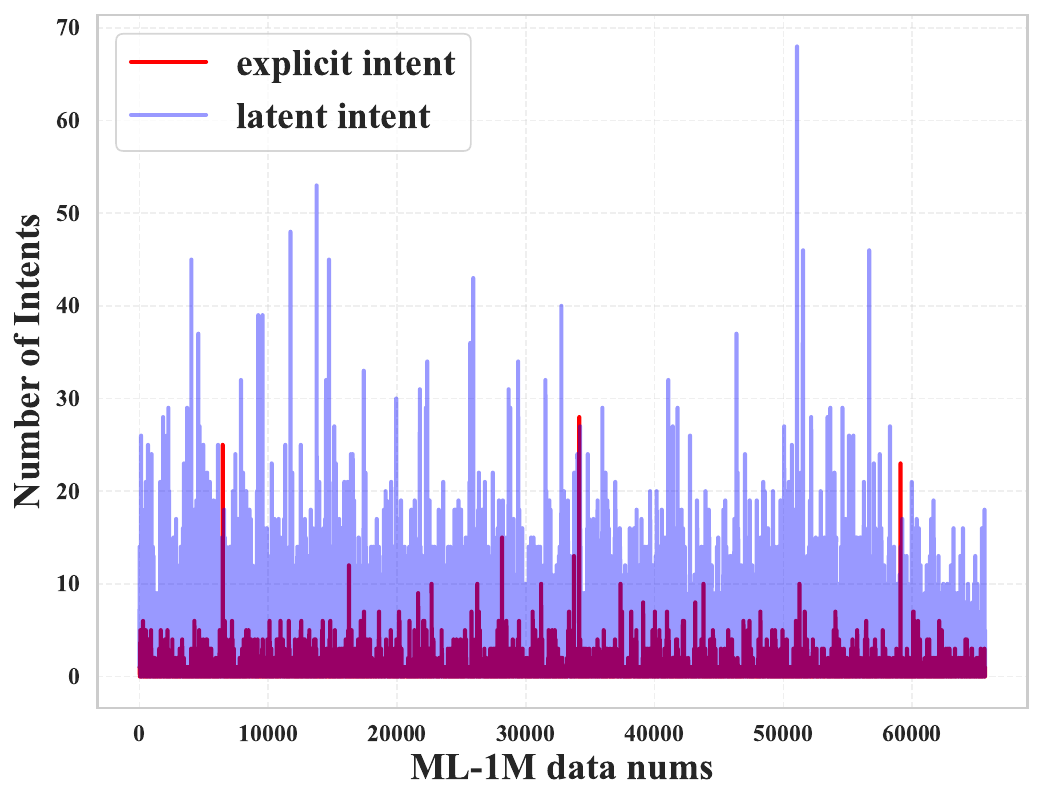}
    \caption{The distribution of explicit and implicit intentions.}\label{fig3}
    \Description{}
\end{figure}

    Building upon the Instruction Prompt, Input Prompt, and the candidate item set constructed in the preceding work, we introduce the Qwen2.5-7B-Instruct\footnote{https://huggingface.co/Qwen/Qwen2.5-7B-Instruct} to infer the user's multi-semantic intentions from the sequence of user interactions. It is important to highlight that the choice of LLM is not fixed; alternative LLMs with enhanced language understanding and reasoning capabilities can also be utilized. In this framework, a question-and-answer format is employed, where the LLM derives and provides the user's intentions at a semantic level based on the given prompts. In particular, the LLM may infer multiple intentions based on the provided prompts. In this work, we formalize the reasoning mechanism as follows:
    \begin{equation}
        \left ( C_{1},C_{2},...,C_{m} \right )=\mathit{F}_\text{Qwen2.5-7B-Instruct}\left ( prompts,s_t^T,\Theta   \right )
    \end{equation}
    Here, $C_{k}$ denotes the $k$-th intent inferred by the LLM, $m$ represents the total number of intents, $s_t^T$ is the user interaction text sequence at time $t$, $\Theta$ refers to the pre-trained parameters of the Qwen2.5-7B-Instruct model, and $\mathit{F}_\text{Qwen2.5-7B-Instruct}$ represents the inference function of the Qwen2.5-7B-Instruct model. Next, we categorize the \( m \) intents inferred by the LLM into two types: explicit intent and latent intent. The categorization rule is as follows:
    \begin{equation}
            C_{k}=\begin{cases}
            \text{explicit intent}, & if \ C_{k} \in s_t^T  \\
            \text{latent intent}, &  otherwise
            \end{cases}
    \end{equation}
    Here, \( k \in \{1, 2, \ldots, m\} \). Based on the aforementioned principle, we divide the set \( (C_1, C_2, \ldots, C_m) \) into two categories: \( \left(C_1^e, C_2^e, \ldots, C_q^e\right) \) and \( \left(C_1^l, C_2^l, \ldots, C_d^l\right) \), where \( C_*^e \) represents explicit intent and \( C_*^l \) represents latent intent. Figure \ref{fig3} illustrates the distribution of both explicit and latent intentions, as identified through Qwen2.5-7B-Instruct reasoning and rule-based classification for the ML-1M dataset. Subsequently, we employ the pre-trained BERT\cite{devlin2018bert} model to encode the two types of intents as follows:
    \begin{equation}
        \left ( \textbf{E}_{1}^e,\textbf{E}_{2}^e,...,\textbf{E}_{q}^e \right )=\xi_{Bert}\left ( C_{1}^e,C_{2}^e,...,C_{q}^e \right )
    \end{equation}
    \begin{equation}
        \left ( \textbf{E}_{1}^l,\textbf{E}_{2}^l,...,\textbf{E}_{d}^l \right )=\xi_{Bert}\left ( C_{1}^l,C_{2}^l,...,C_{d}^l \right )
    \end{equation}
    In this context, \( \mathbf{E}_*^e \) and \( \mathbf{E}_*^l \) represent the embeddings of explicit intent and latent intent, respectively. To capture the overall features of these two types of intent, we apply the average pooling operation to aggregate \( \mathbf{E}_{i}^e \)($i\in \left\{ 1,2,...,q\right\}$) and \( \mathbf{E}_{k}^l \)($k\in \left\{ 1,2,...,d\right\}$), yielding the explicit intent feature \( \mathbf{E}_e = \frac{1}{q} \sum_{i=1}^{q} \mathbf{E}_i^e \) and the latent intent feature \( \mathbf{E}_l = \frac{1}{d} \sum_{k=1}^{d} \mathbf{E}_k^l \), respectively.

\subsection{Intent Alignment and Training}
    The explicit intent features \( \mathbf{E}_e \in \mathbb{R}^{d \times v} \) and latent intent features \( \mathbf{E}_l \in \mathbb{R}^{d \times v} \), as learned above, are derived from the user interaction text sequences. In the GNNs-based method, the input to the model is the product ID sequence from user interactions, which is represented as graph-structured data. GNNs are employed to learn the information flow between the nodes in the graph, ultimately yielding the final structural intent $\bar{\mathbf{G}} \in \mathbb{R}^{d \times d}$. Specifically, given the interaction sequence $s_t = (v_1, v_2, \dots, v_l)$ of a user at time $t$, the structural intent learned by the GNN can be expressed as:
    \begin{equation}
        \bar{\mathbf{G}}=\text{GNN-model}(s_t)
    \end{equation}
    To unify \( \mathbf{E}_e \), \( \mathbf{E}_l \), and \( \bar{\mathbf{G}} \) in the embedding space, we apply a linear transformation to \( \mathbf{E}_e \) and \( \mathbf{E}_l \) to obtain \( \bar{\mathbf{E}}_e = \mathbf{w} \mathbf{E}_e + \mathbf{b} \) and \( \bar{\mathbf{E}}_l = \mathbf{w} \mathbf{E}_l + \mathbf{b} \), where \( \bar{\mathbf{E}}_e \in \mathbb{R}^{d \times d} \) and \( \bar{\mathbf{E}}_l \in \mathbb{R}^{d \times d} \). In particular, the GNN-based model is not fixed and can include methods such as SR-GNN \cite{wu2019session}, GCE-GNN \cite{wang2020global}, DHCN \cite{xia2021self}, and others. Next, a linear transformation is applied to aggregate the semantic intent and structural intent, yielding the final session representation \( \mathbf{S} \), which can be expressed as follows:
    \begin{equation}
        \mathbf{S}=\mathbf{W}\left [ \bar{\mathbf{G}},\bar{\mathbf{E}}_e,\bar{\mathbf{E}}_l \right ]
    \end{equation}
    Here, $\mathbf{W} \in \mathbb{R}^{d \times 2d}$ compresses the two combined embedding vectors in the latent space $\mathbb{R}^d$. To effectively align the semantic intent learned by LLMs and the structural intent learned by GNNs, we introduce the Kullback-Leibler (KL)\cite{kullback1951information} divergence to measure the difference between the two distributions. The specific calculation is given by:
    \begin{equation}
         \mathcal{L}_e = \text{KL}(\bar{\mathbf{G}} \parallel \bar{\mathbf{E}}_e) = \sum_{i=1}^{n} \widetilde{\textbf{G}}(i) \log \frac{\widetilde{\textbf{G}}(i)}{\widetilde{\mathbf{E}}_e(i)}
    \end{equation}
    \begin{equation}
        \mathcal{L}_l = \text{KL}(\bar{\mathbf{G}} \parallel \bar{\mathbf{E}}_l) = \sum_{i=1}^{n} \widetilde{\textbf{G}}(i) \log \frac{\widetilde{\textbf{G}}(i)}{\widetilde{\mathbf{E}}_l(i)}
    \end{equation}
    \begin{equation}
        \mathcal{L}_{info} = \alpha\mathcal{L}_e + \beta\mathcal{L}_l
    \end{equation}
    Where \( \widetilde{\mathbf{G}}(k) = \frac{e^{G(k)}}{\sum_{j=1}^{n} e^{G(j)}} \), \( \widetilde{\mathbf{E}}_e(k) = \frac{e^{E_e(k)}}{\sum_{j=1}^{n} e^{E_e(j)}} \), and \( \widetilde{\mathbf{E}}_l(k) = \frac{e^{E_l(k)}}{\sum_{j=1}^{n} e^{E_l(j)}} \) represent the probability distributions of the structural intent \( \bar{\mathbf{G}} \), explicit intent features \( \bar{\mathbf{E}}_e \), and latent intent features \( \bar{\mathbf{E}}_l \), respectively. $\alpha$ and $\beta$ are hyperparameters that represent the weighting coefficients for the explicit intent alignment loss $\mathcal{L}_e$ and the latent intent alignment loss $\mathcal{L}_l$, respectively. 

\subsection{Prediction}
    Based on the session representation \( \mathbf{S} \) obtained above, we apply the softmax function to compute the model's predicted value \( \hat{y} \).
    \begin{equation}
        {\hat{y}}_i=softmax(\mathbf{S}^T\mathbf{v}_i)
    \end{equation}
    Here, \( \hat{y}_i \) represents the probability of the user's next click on the candidate items. The cross-entropy loss function is then applied to compute the loss between the predicted value and the true value:
    \begin{equation}
        \mathcal{L}_r=-\sum_{i=1}^{n}{y_ilog({\hat{y}}_i)}+(1-y_i)log(1-{\hat{y}}_i)
    \end{equation}
    Here, \( y \) represents the ground-truth value. The final joint loss function combines the recommendation loss function \( \mathcal{L}_r \) of the main task and the alignment loss function \( \mathcal{L}_k \) of the auxiliary task, which can be expressed as follows:
    \begin{equation}
        \mathcal{L}=\mathcal{L}_r+\sigma\mathcal{L}_{info}
    \end{equation}
    \( \sigma \) is a hyperparameter representing the ratio coefficient of the auxiliary task. 
    
    As shown in Algorithm \ref{alg:training}, we present the training process of our proposed framework, LLM-DMsRec, through pseudocode.

\begin{algorithm}
    \caption{\textbf{Training Procedure in LLM-DMsRec}}
    \label{alg:training}
    \small 
    \begin{algorithmic}[1]
        \Require Base model $\mathcal{R}$, implicit feedback $\mathcal{X}$, session sequence $S$ of the user interaction, learning rate $\eta$
        \Ensure Trained model parameters $\Theta$
        \State Pre-trained GNN to obtain candidate item set $T$.
        \State LLM infers users' multiple semantic intentions based on $S$ and $T$.
        \State \textbf{repeat}
        \State \quad Uniformly sample batch data $\mathcal{B} = \{(S \in \mathcal{X}, \mathbf{E}_*^e, \mathbf{E}_*^l)\}$;
        \State \quad Learning graph structure intent $\bar{\mathbf{G}}$ with $\mathcal{R}$;
        \State \quad Infer item representation $e_{v}$ with $\mathcal{R}$;
        \State \quad Calculate model optimization objective $\mathcal{L_R}$ based on $\mathcal{B}$;
        \State \quad Calculate $L_{info}$ \textit{w.r.t.} Eq (10 \& 11 \& 12) for $\mathbf{E}_*^e$, $\mathbf{E}_*^l$ and $\bar{\mathbf{G}}$;
        \State \quad $\mathcal{L} = \mathcal{L_R} + \sigma L_{info}$;
        \State \quad $\Theta \gets \Theta - \eta \nabla_{\Theta} \mathcal{L}$;
        \State \textbf{until} convergence;
    \end{algorithmic}
\end{algorithm}

\section{Experiment}
    To validate the effectiveness of our proposed model, \textbf{LLM-DMsRec}, its performance is assessed by addressing the following key research questions:
    \begin{itemize}[leftmargin=*]
        \item \textbf{RQ1} How does the performance of the LLM-DMsRec framework, when integrated into the GNN model, compare with that of the traditional GNN model?
        \item \textbf{RQ2} Can the semantic intent inferred from LLMs contribute to improving recommendation performance? Additionally, is it crucial to differentiate between explicit and implicit intent? Finally, is the incorporation of an alignment mechanism essential?
        \item \textbf{RQ3} How do various alignment strategies influence performance?
        \item \textbf{RQ4} How do different hyperparameter configurations, such as the auxiliary scaling factor $\sigma$ impact the model's performance?
    \end{itemize}



\begin{table}[htbp]
    \centering
    \caption{Statistics of the utilized datasets. }
    \begin{tabular}{c|ccccc}
    \toprule
    \textbf{Datasets} & \textbf{\#Train} & \textbf{\#Test} & \textbf{\#Clicks} & \textbf{\#Items} & \textbf{Avg.len}\\
    \hline
    Beauty & 158,717 & 17,422 & 198,502 & 12,101 & 8.66\\
    ML-1M & 69,906 & 7,768 & 999,611 & 3,416 & 12.79\\
    \bottomrule
  \end{tabular}
  \label{table1}
\end{table}


\subsection{Experiments settings}
\subsubsection{Datasets and preprocessing.}
    In this experiment, we use two representative real-world datasets: Beauty and MovieLens-1M (ML-1M)\footnote{https://grouplens.org/datasets/movielens/}. Please refer to Table \ref{table1} for a summary of the dataset statistics. Following the approach of previous studies\cite{wu2019session,wang2020global,li2017neural}, we preprocess the data by removing sessions with fewer than one interaction and those that appear fewer than five times across all sessions. The detailed descriptions of these datasets are as follows:
    \begin{itemize}[leftmargin=*]
        \item \textbf{Beauty} used in this study is an e-commerce dataset derived from user reviews provided by Amazon\footnote{ https://jmcauley.ucsd.edu/data/amazon/links.html}, focusing specifically on beauty products. To enhance the data, we adopt the prefix subsequence segmentation method described in \cite{wu2019session, wang2020global, xia2021self}. Given a session sequence \( S = [v_1, v_2, v_3, v_4] \), its prefix subsequences are:\([v_1], [v_1, v_2], [v_1, v_2, v_3]\). In particular, we retain only those subsequences with a length greater than 1 and add them to the dataset.
        \item \textbf{ML-1M} contains more than 1 million ratings of nearly 4,000 movies by more than 6,000 users. To facilitate the analysis of the problem, we partition the data by timestamps, with a 5-minute interval serving as the time segmentation point within a 24-hour period. If the user does not engage in any new rating behavior within the designated time period, or if the time exceeds 24 hours, the current session is considered to have concluded, and a new session is initiated.
    \end{itemize}
\subsubsection{Implementation Details.}
To ensure fairness in the experiment, we set the batch size for all models to 100, the learning rate to 0.001, which is decayed by a factor of 0.1 every three epochs, and the L$_2$ regularization term to $10^{-5}$. The encoding dimension was set to 100, and Adam\cite{diederik2014adam} was used as the optimizer. All experiments were conducted on an NVIDIA A800 80GB GPU. This framework sets the number of candidate item set to 50. For the other parameters in the backbone models, we follow the optimal settings reported in their respective papers.
\subsubsection{Evaluation metrices.}
To facilitate comparison with the baseline model, this experiment employs two commonly used evaluation metrics in SBR, namely Precision (P) and Mean Reciprocal Rank (MRR). Specifically, to assess the recommendation performance across varying numbers of recommended items, we utilize P@K and MRR@K, where K represents the number of items recommended.


\subsubsection{BackBone and Baselines}
This paper uses five classic GNN-based SBR models as the backbone, namely SR-GNN\cite{wu2019session}, TAGNN\cite{yu2020tagnn}, SHARE\cite{wang2021session}, GCE-GNN\cite{wang2020global} and MSGIFSR\cite{guo2022learning}. At the same time, we select two models that capture users' multi-level intents as baseline models: Atten-Mixer\cite{zhang2023efficiently} and MiaSRec\cite{choi2024multi}. The details of the compared models are described as follows:
\begin{itemize}[leftmargin=*]
    \item \textbf{SR-GNN}\footnote{https://github.com/CRIPAC-DIG/SR-GNN} is the first method to model session sequences as graph-structured data and leverage GNN to capture the intricate transition relationships between items.
    \item \textbf{TAGNN} \footnote{https://github.com/CRIPAC-DIG/TAGNN} introduces a target-attention graph neural network model that adaptively activates distinct user interests.
    \item \textbf{SHARE}\footnote{https://github.com/wangjlgz/Hypergraph-Session-Recommendation} models the item correlations and learns a dynamic session representation using hypergraph attention.
    \item \textbf{GCE-GNN}\footnote{https://github.com/CCIIPLab/GCE-GNN} infers user intent by learning item embeddings from both the session graph and the global graph.
    \item \textbf{MSGIFSR}\footnote{https://github.com/SpaceLearner/SessionRec-pytorch} enhances session-based recommendation performance by effectively capturing user intents at multiple granularities.
    \item \textbf{Atten-Mixer}\footnote{https://github.com/Peiyance/Atten-Mixer-torch} captures user's multi-level intentions by eliminating redundant GNN propagation and integrating multi-view readout modules with inductive biases.
    \item \textbf{MiaSRec}\footnote{https://github.com/jin530/MiaSRec} introduces frequency embedding and multi-session representations to capture users' diverse intentions and dynamically select key intents.
\end{itemize}


\subsection{Overall performance comparision(RQ1)}
    \textbf{Overall comparison.} Table \ref{tab2} shows the overall performance of our proposed LLM-DMsRec framework on the Beauty and ML-1M datasets, respectively. As shown in the table, integrating LLM into the GNN-based SBR method leads to a substantial improvement in the performance of the recommendation system. Notably, LLM-DMsRec (GCE-GNN) enhances the MRR@5 metric by 42.56\% and 65.33\% on the Beauty and ML-1M datasets, respectively. This indicates that the semantic information inferred by the LLM can effectively enrich the user's intent, which is crucial for enhancing the recommendation performance. The table also shows that, compared to the baseline models, LLM-DMsRec (MSGIFSR) achieves the highest performance. Additionally, while GCE-GNN exhibits lower overall performance than MiaSRec, LLM-DMsRec (GCE-GNN) outperforms MiaSRec, further validating the effectiveness of the proposed LLM-DMsRec framework. 
    
    \textbf{Comparison between MRR@K and P@K.} As shown in the table, the performance of the proposed LLM-DMsRec framework is generally superior on the MRR@K evaluation metric compared to P@K. For instance, LLM-DMsRec (GCE-GNN) achieves improvements of 42.56\%, 34.25\%, and 30.97\% on MRR@5, MRR@10, and MRR@20, respectively, on the Beauty dataset, while only showing improvements of 11.97\%, 0.21\%, and a slight decline of 2.91\% on P@5, P@10, and P@20, respectively. This suggests that LLM is more effective in capturing user intentions and preferences by leveraging rich semantic information and reasoning capabilities. Additionally, LLM's ability to infer implicit information from session sequences enhances the recognition of rare items, thereby significantly improving the ranking of relevant items in the recommendation results. Notably, the table also shows that LLM-DMsRec (GCE-GNN) exhibits a decrease of 2.91\% on the Beauty dataset and 4.41\% on the ML-1M dataset in the P@20 metric. This indicates that LLM may introduce irrelevant or low-quality recommendations during the reasoning process, resulting in redundant or inaccurate items in longer recommendation lists. 

    \textbf{K value comparison.} As observed in the table, the LLM-DMsRec framework demonstrates a higher proportion of improvement in recommendation performance for smaller K values. For instance, on the ML-1M dataset, LLM-DMsRec (SR-GNN) achieves performance gains of 13.36\%, 4.66\%, and 5.03\% on P@5, P@10, and P@20, respectively, and improves by 23.20\%, 18.69\%, and 17.06\% on MRR@5, MRR@10, and MRR@20, respectively. These results highlight that the semantic information inferred by LLMs can significantly enhance ranking and accuracy, particularly for shorter recommendation lists. However, the semantic inference by LLMs may introduce noise, and its impact on recommendation performance becomes more pronounced as the K value increases.
    

    
    
    

\begin{table*}[t]
    \centering
    \caption{A Table with overall performance on Beauty and ML-1M. Improve. indicates the performance improvement compared to the backbone model. The best performing model is marked in \textbf{bold}, and the second best model is \underline{underlined}.}
    \resizebox{\textwidth}{!}{
    \begin{tabular}{ccccccc|cccccc}
    \toprule
    \multicolumn{1}{c}{\textbf{Dataset}} & \multicolumn{6}{c}{\textbf{Beauty}} & \multicolumn{6}{c}{\textbf{ML-1M}} \\  
    \cmidrule(lr){1-1} \cmidrule(lr){2-7} \cmidrule(lr){8-13}
    \textbf{Model} & \textbf{P@5} & \textbf{P@10} & \textbf{P@20} & \textbf{MRR@5} & \textbf{MRR@10} & \textbf{MRR@20} & \textbf{P@5} & \textbf{P@10} & \textbf{P@20} & \textbf{MRR@5} & \textbf{MRR@10} & \textbf{MRR@20} \\
    \midrule
   Atten-Mixer & 10.0447 & 13.4312 & 17.3171 & 6.4574 & 6.9060 & 7.1676 & 5.8538 & 8.5967 & 12.0254 & 3.6218 & 3.9852 & 4.2163\\
   MiaSRec & 9.0862 & 12.8630 & 17.1565 & 5.7439 & 6.2439 & 6.5393 & 5.0677 & 7.7772 & 12.3432 & 2.8990 & 3.2541 & 3.5663\\
   \hline\hline
   SR-GNN & 7.3872 & 10.9230 & 15.6526 & 4.1421 & 4.5971 & 4.9231 & 3.6294 & 6.1047 & 9.9682 & 1.9058 & 2.2262 & 2.4927\\
   \textbf{LLM-DMsRec(SR-GNN)} & 8.1621 & 11.8815 & 16.6743 & 4.5511 & 5.0447 & 5.3578 & 4.1144 & 6.3890 & 10.4700 & 2.3479 & 2.6422 & 2.9179\\
   SR-GNN Improve. & 10.49\%\textsuperscript{$\uparrow$}  & 8.78\%\textsuperscript{$\uparrow$} & 6.53\%\textsuperscript{$\uparrow$} & 9.87\%\textsuperscript{$\uparrow$} & 9.74\%\textsuperscript{$\uparrow$} & 8.83\%\textsuperscript{$\uparrow$}  & 13.36\%\textsuperscript{$\uparrow$} & 4.66\%\textsuperscript{$\uparrow$} & 5.03\%\textsuperscript{$\uparrow$} & 23.20\%\textsuperscript{$\uparrow$} & 18.69\%\textsuperscript{$\uparrow$} & 17.06\%\textsuperscript{$\uparrow$}\\
   \hline
   TAGNN & 7.1634 & 10.7623 & 15.2566 & 4.0242 & 4.4925 & 4.8059 & 3.2781 & 5.4022 & 8.8476 & 1.7383 & 2.0041 & 2.2366\\
   \textbf{LLM-DMsRec(TAGNN)} & 8.1334 & 12.0021 & 16.9097 & 4.5758 & 5.0829 & 5.4180 & 4.1980 & 6.6734 & 10.4867 & 2.1994 & 2.5221 & 2.7784\\
   TAGNN Improve. & 13.54\%\textsuperscript{$\uparrow$} & 11.52\%\textsuperscript{$\uparrow$} & 10.84\%\textsuperscript{$\uparrow$} & 13.71\%\textsuperscript{$\uparrow$} & 13.14\%\textsuperscript{$\uparrow$} & 12.74\%\textsuperscript{$\uparrow$} & 28.06\%\textsuperscript{$\uparrow$} & 23.53\%\textsuperscript{$\uparrow$} & 18.53\%\textsuperscript{$\uparrow$} & 26.53\%\textsuperscript{$\uparrow$} & 25.83\%\textsuperscript{$\uparrow$} & 24.22\%\textsuperscript{$\uparrow$}\\
   \hline
   SHARE & 8.1047 & 11.6749 & 16.0946 & 5.0588 & 5.5317 & 5.8334 & 4.3653 & 6.9577 & 10.6874 & 2.3950 & 2.7469 & 3.0011\\
   \textbf{LLM-DMsRec(SHARE)} & 8.5696 & 12.2144 & 16.9900 & 5.0855 & 5.5677 & 5.8929 & 4.6496 & 7.2755 & 11.0721 & 2.4224 & 2.7601 & 3.0108\\
   SHARE Improve. & 5.74\%\textsuperscript{$\uparrow$} & 4.62\%\textsuperscript{$\uparrow$} & 5.56\%\textsuperscript{$\uparrow$} & 0.53\%\textsuperscript{$\uparrow$} & 0.65\%\textsuperscript{$\uparrow$} & 1.02\%\textsuperscript{$\uparrow$} & 6.51\%\textsuperscript{$\uparrow$} & 4.57\%\textsuperscript{$\uparrow$} & 3.60\%\textsuperscript{$\uparrow$} & 1.14\%\textsuperscript{$\uparrow$} & 0.48\%\textsuperscript{$\uparrow$} & 0.32\%\textsuperscript{$\uparrow$}\\
   \hline
    GCE-GNN & 8.8681 & 13.4485 & 18.8497 & 4.3143 & 4.9262 & 5.3011 & 4.6663 & 8.1954 & \underline{13.2631} & 1.9797 & 2.4447 & 2.7717\\
   \textbf{LLM-DMsRec(GCE-GNN)} & 9.9300 & 13.4772 & 18.3159 & 6.1504 & 6.6136 & 6.9427 & 5.7702 & 8.2623 & 12.6777 & 3.2731 & 3.6004 & 3.8917\\
    GCE-GNN Improve. & 11.97\%\textsuperscript{$\uparrow$} & 0.21\%\textsuperscript{$\uparrow$} & -2.91\%\textsuperscript{$\downarrow$} & 42.56\%\textsuperscript{$\uparrow$} & 34.25\%\textsuperscript{$\uparrow$} & 30.97\%\textsuperscript{$\uparrow$} & 23.66\%\textsuperscript{$\uparrow$} & 0.82\%\textsuperscript{$\uparrow$} & -4.41\%\textsuperscript{$\downarrow$} & 65.33\%\textsuperscript{$\uparrow$} & 47.27\%\textsuperscript{$\uparrow$} & 40.41\%\textsuperscript{$\uparrow$}\\
   \hline
   MSGIFSR & \underline{10.7343} & \underline{14.5933} & \underline{19.3085} & \underline{6.9131} & \underline{7.4275} & \underline{7.7326} & \underline{5.9876} & \underline{9.0316} & 12.6944 & \underline{3.6985} & \underline{4.1136} & \underline{4.3636}\\
   \textbf{LLM-DMsRec(MSGIFSR)} & \textbf{11.2611} & \textbf{14.8502} & \textbf{19.6971} & \textbf{7.1038} & \textbf{7.5943} & \textbf{7.9316} & \textbf{6.2051} & \textbf{9.2658} & \textbf{13.3634} & \textbf{3.8574} & \textbf{4.2681} & \textbf{4.5248}\\
   MSGIFSR Improve. & 4.91\%\textsuperscript{$\uparrow$} & 1.76\%\textsuperscript{$\uparrow$} & 2.01\%\textsuperscript{$\uparrow$} & 2.76\%\textsuperscript{$\uparrow$} & 2.24\%\textsuperscript{$\uparrow$} & 2.57\%\textsuperscript{$\uparrow$} & 3.63\%\textsuperscript{$\uparrow$} & 2.59\%\textsuperscript{$\uparrow$} & 5.27\%\textsuperscript{$\uparrow$} & 4.30\%\textsuperscript{$\uparrow$} & 3.75\%\textsuperscript{$\uparrow$} & 3.69\%\textsuperscript{$\uparrow$}\\
   \bottomrule
  \end{tabular}
    }
  \label{tab2}
\end{table*}

\subsection{Ablation studies}
    \label{ablation section}
    \subsubsection{Different modules of LLM-DMsRec(RQ2).}
    To evaluate the impact of different modules within the LLM-DMsRec framework on experimental performance, the model is divided into four variants: w/o Semantic, w/o KL, w/o Latent-Intent, and w/o Explicit-Intent. The details of these four modules are described as follows:
    \begin{itemize}[leftmargin=*]
        \item \textbf{w/o Semantic} removes the semantic information module, retaining only the original GNN model and utilizing solely the ID information within the session sequence.
        \item \textbf{w/o KL} variant eliminates the alignment module and excludes the use of KL divergence to align semantic and structural information.
        \item \textbf{w/o Latent-Intent} variant modifies the GNN by fusing only explicit intent while excluding latent intent during the integration of semantic information.
        \item \textbf{w/o Explicit-Intent} variant modifies the GNN to fuse only latent intent during the integration of semantic information, while excluding explicit intent.
    \end{itemize}

    Table \ref{tab3} presents the performance of different modules of LLM-DMsRec (GCE-GNN) on the Beauty dataset, evaluated using P@K and MRR@K. As shown in the table, LLM-DMsRec (GCE-GNN) consistently outperforms the other modules. The results for the w/o Semantic variant demonstrate the importance of semantic information in enhancing recommendation performance. However, the semantic intent inferred by LLM may contain noise, leading to a slight decline in performance for longer recommendation lists. The w/o KL results highlight the significance of the alignment module, as KL divergence effectively aligns the semantic information learned by LLM with the structural information learned by GNN, which reside in distinct feature spaces. For the w/o Latent-Intent and w/o Explicit-Intent variants, the results indicate that both explicit and latent user intent contribute to recommendation performance, emphasizing the necessity of separately modeling explicit and latent intent inferred by LLM.

    

\begin{table}[t]
    \centering
    \caption{Performance comparison of different modules on the Beauty}
    \small  
    \resizebox{0.85\columnwidth}{!} {  
    \begin{tabular}{ccccccc}
    \toprule
    \multicolumn{1}{c}{\textbf{Dataset}} & \multicolumn{6}{c}{\textbf{Beauty}} \\
    \cmidrule(lr){1-1} \cmidrule(lr){2-7}
    \textbf{Model} & \textbf{P@5} & \textbf{P@10} & \textbf{P@20} & \textbf{MRR@5} & \textbf{MRR@10} & \textbf{MRR@20} \\
    \midrule
    w/o Semantic & 8.8681 & 13.4485 & \textbf{18.8497} & 4.3143 & 4.9262 & 5.3011\\
    w/o KL & 9.069 & 12.4785 & 16.9211 & 5.5534 & 5.9783 & 6.2692\\
    w/o Latent-Intent & 9.6659 & 12.8401 & 17.3229 & 5.9907 & 6.4099 & 6.7036\\
    w/o Explicit-Intent & 9.4937 & 12.9664 & 17.3459 & 5.9649 & 6.4103 & 6.7082\\
    \hline
    \textbf{LLM-DMsRec(GEC-GNN)} & \textbf{9.93} & \textbf{13.4772} & 18.3159 & \textbf{6.1504} & \textbf{6.6136} & \textbf{6.9427}\\
    \bottomrule
    \end{tabular}
    }
    \label{tab3}
\end{table}

\begin{figure}[t]
    \centering
    \includegraphics[height=8cm,width=10cm]{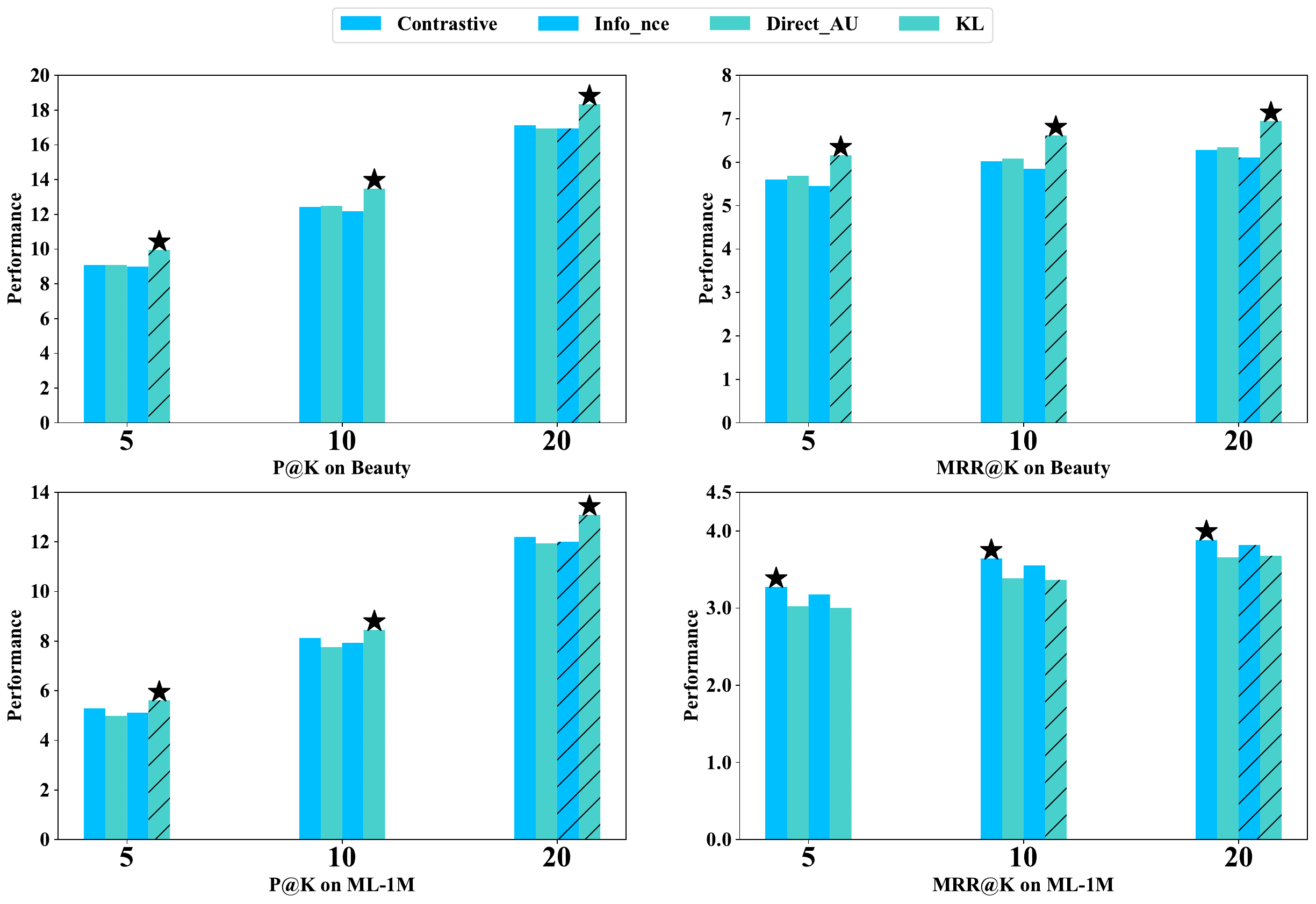}
    \caption{Impact of different alignment mechanisms on performance}\label{fig4}
    \Description{}
\end{figure}

 \subsubsection{Alignment mechanisms(RQ3).}
    The semantic intent (text-based) learned by the LLM and the structural intent (ID-based) learned by the GNN reside in distinct feature spaces. Effectively aggregating these two types of information is crucial for enhancing recommendation performance. To investigate the impact of different intent alignment methods on recommendation performance, we designed three alignment mechanisms for the experiment: contrastive learning, InfoNCE\cite{oord2018representation}, and DirectAU\cite{wang2022towards}.

    Figure \ref{fig4} shows the performance of different alignment methods for LLM-DMsRec (GCE-GNN) on Beauty and ML-1M datasets. As observed, the KL alignment mechanism outperforms the other three methods, indicating that KL divergence effectively minimizes the discrepancy between the two distributions. This facilitates the complementary integration of semantic and structural information, thereby improving recommendation performance. However, for the ML-1M dataset and the MRR@K metric, the contrastive learning-based alignment mechanism performs better than KL divergence. We speculate that the noise present in the semantic information makes it challenging for KL divergence to effectively align the two distributions. In contrast, contrastive learning utilizes the structure of positive and negative sample pairs, which helps to mitigate the influence of noise.

\begin{figure}[t]
    \centering
    \includegraphics[width=10cm]
    {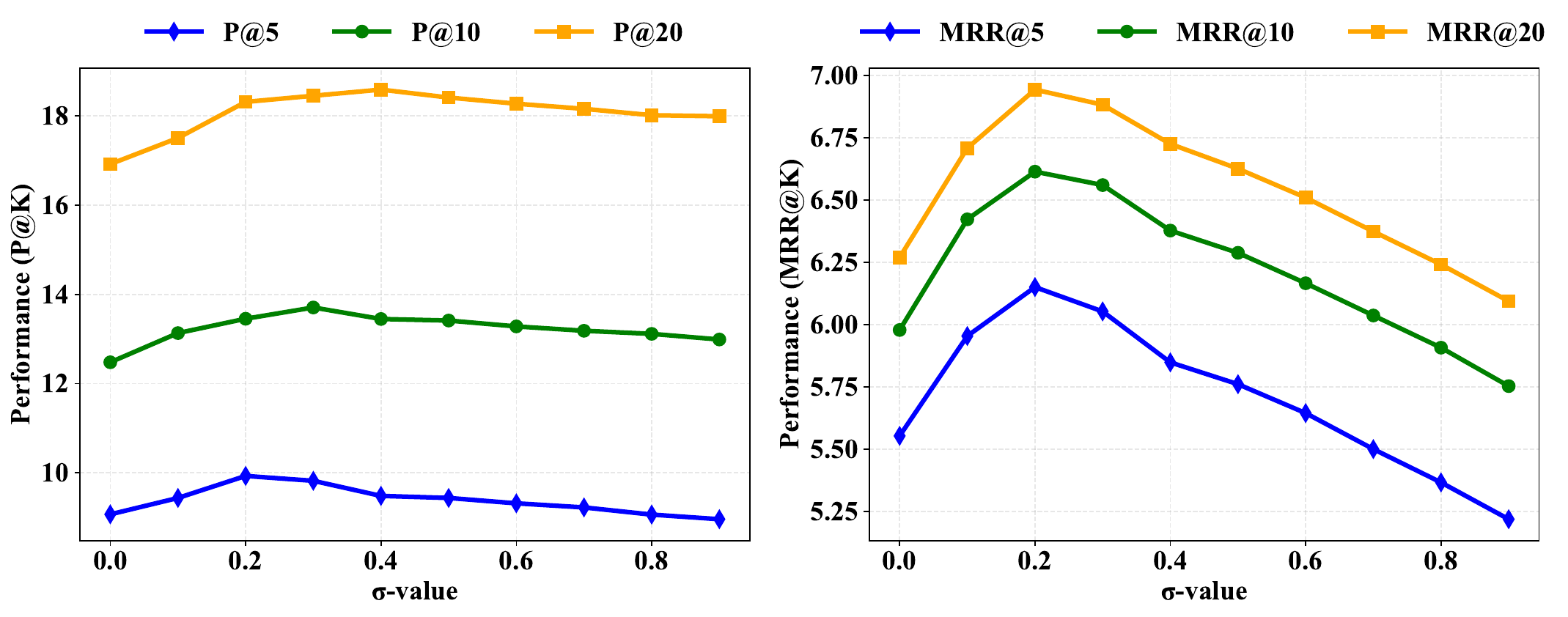}
    \caption{Impact of different $\sigma$ on performance}\label{fig5}
    \Description{}
\end{figure}
    
\begin{figure}[t]
    \centering
    \includegraphics[width=10cm]{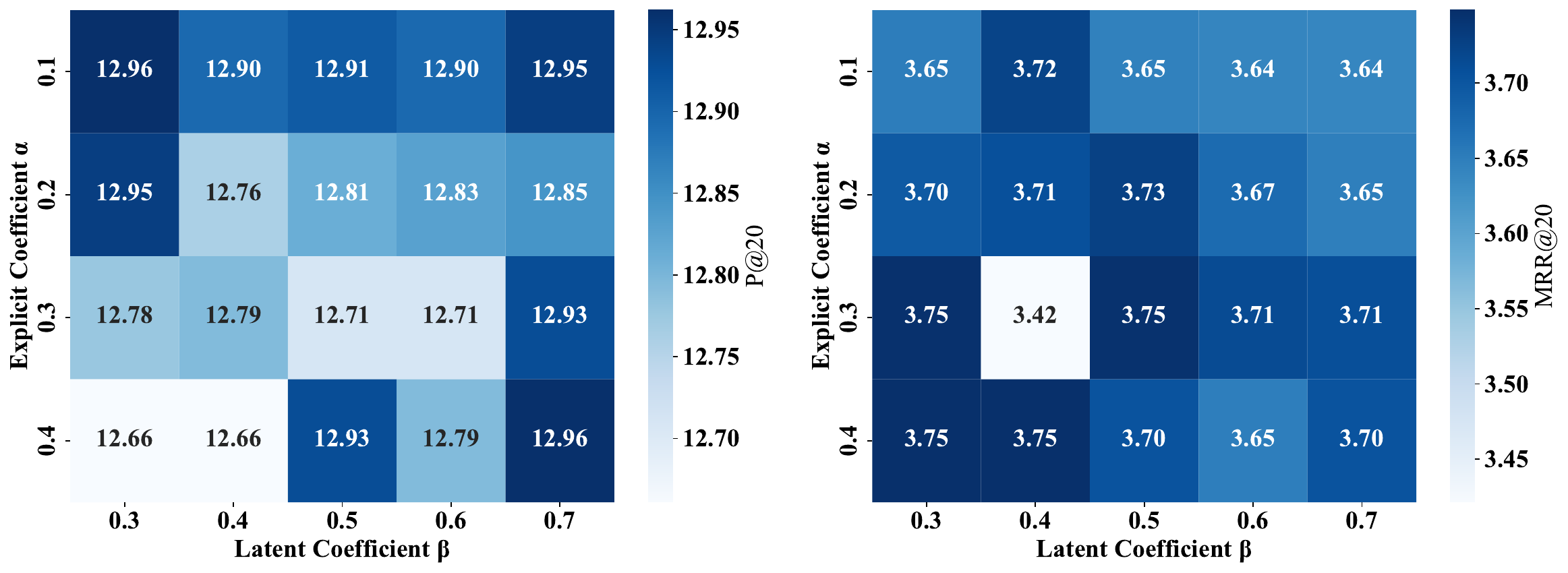}
    \caption{Impact of explicit intent coefficient $\alpha$ and latent intent coefficient $\beta$ on performance.} \label{fig6}
    \Description{}
\end{figure}


\subsubsection{Hyperparameters(RQ4)}
    The auxiliary factor \( \sigma \) controls the contribution of the auxiliary task to the total loss. To analyze its impact on performance, we vary \( \sigma \) from 0.0 to 0.9. Figure \ref{fig5} presents the performance of different \( \sigma \) values for the LLM-DMsRec (GCE-GNN) model, evaluated using the P@K and MRR@K metrics on the Beauty dataset. As shown, the recommendation performance on MRR@K reaches its peak when \( \sigma = 0.2 \). 
    Interestingly, as \( \sigma \) increases, the overall recommendation performance initially improves, peaking at an optimal value, and then begins to decline. This trend indicates that a moderate \( \sigma \) enhances the positive influence of the auxiliary task on the main task, leading to improved performance. However, exceeding the optimal value results in the auxiliary task overly interfering with the main task, introducing noise and reducing performance. 

    Figure \ref{fig6} illustrates the impact of the explicit intent coefficient $\alpha$ and the latent intent coefficient $\beta$ on the performance of the LLM-DMsRec (GCE-GNN) model, evaluated on the ML-1M dataset using the metrics P@20 and MRR@20. The results demonstrate that varying the values of $\alpha$ and $\beta$ has a notable effect on model performance. Specifically, our results demonstrate that for P@20, the optimal performance is achieved when $\alpha = 0.1$. In the case of MRR@20, performance improves with smaller values of $\beta$ and larger values of $\alpha$.

\section{Conclusion}
    This paper introduces \textbf{LLM-DMsRec}, a novel framework that integrates LLM-driven multi-semantic intent into graph models for session-based recommendation. This framework first generates a set of candidate items using a pre-trained GNN model, which serves as a knowledge base to alleviate hallucinations during LLM reasoning. Next, Qwen2.5-7B-Instruct is employed to infer the user's multi-semantic intent. Specifically, the user's intent is classified into explicit and latent intents based on predefined rules, with a pre-trained BERT model employed to encode both types of intent. Finally, the encoded intent representations are integrated into the GNN model for joint training, with KL divergence employed to align both semantic and structural information. Extensive experiments on the Beauty and ML-1M datasets validate that the proposed LLM-DMsRec framework effectively enhances the recommendation performance of GNN models.
    
    In future work, we aim to further investigate the application of LLMs in session-based recommendation systems. Specifically, we will focus on exploring more effective approaches to accurately infer user intent leveraging the capabilities of LLMs.


\bibliographystyle{ACM-Reference-Format}
\bibliography{reference}


\begin{thebibliography}{45}


\ifx \showCODEN    \undefined \def \showCODEN     #1{\unskip}     \fi
\ifx \showDOI      \undefined \def \showDOI       #1{#1}\fi
\ifx \showISBNx    \undefined \def \showISBNx     #1{\unskip}     \fi
\ifx \showISBNxiii \undefined \def \showISBNxiii  #1{\unskip}     \fi
\ifx \showISSN     \undefined \def \showISSN      #1{\unskip}     \fi
\ifx \showLCCN     \undefined \def \showLCCN      #1{\unskip}     \fi
\ifx \shownote     \undefined \def \shownote      #1{#1}          \fi
\ifx \showarticletitle \undefined \def \showarticletitle #1{#1}   \fi
\ifx \showURL      \undefined \def \showURL       {\relax}        \fi
\providecommand\bibfield[2]{#2}
\providecommand\bibinfo[2]{#2}
\providecommand\natexlab[1]{#1}
\providecommand\showeprint[2][]{arXiv:#2}

\bibitem[Achiam et~al\mbox{.}(2023)]%
        {achiam2023gpt}
\bibfield{author}{\bibinfo{person}{Josh Achiam}, \bibinfo{person}{Steven Adler}, \bibinfo{person}{Sandhini Agarwal}, \bibinfo{person}{Lama Ahmad}, \bibinfo{person}{Ilge Akkaya}, \bibinfo{person}{Florencia~Leoni Aleman}, \bibinfo{person}{Diogo Almeida}, \bibinfo{person}{Janko Altenschmidt}, \bibinfo{person}{Sam Altman}, \bibinfo{person}{Shyamal Anadkat}, {et~al\mbox{.}}} \bibinfo{year}{2023}\natexlab{}.
\newblock \showarticletitle{Gpt-4 technical report}.
\newblock \bibinfo{journal}{\emph{arXiv preprint arXiv:2303.08774}} (\bibinfo{year}{2023}).
\newblock


\bibitem[Anyosa et~al\mbox{.}(2018)]%
        {anyosa2018incremental}
\bibfield{author}{\bibinfo{person}{Susan~C Anyosa}, \bibinfo{person}{Jo{\~a}o Vinagre}, {and} \bibinfo{person}{Al{\'\i}pio~M Jorge}.} \bibinfo{year}{2018}\natexlab{}.
\newblock \showarticletitle{Incremental matrix co-factorization for recommender systems with implicit feedback}. In \bibinfo{booktitle}{\emph{Companion Proceedings of the The Web Conference 2018}}. \bibinfo{pages}{1413--1418}.
\newblock


\bibitem[Bai et~al\mbox{.}(2023)]%
        {bai2023qwen}
\bibfield{author}{\bibinfo{person}{Jinze Bai}, \bibinfo{person}{Shuai Bai}, \bibinfo{person}{Yunfei Chu}, \bibinfo{person}{Zeyu Cui}, \bibinfo{person}{Kai Dang}, \bibinfo{person}{Xiaodong Deng}, \bibinfo{person}{Yang Fan}, \bibinfo{person}{Wenbin Ge}, \bibinfo{person}{Yu Han}, \bibinfo{person}{Fei Huang}, {et~al\mbox{.}}} \bibinfo{year}{2023}\natexlab{}.
\newblock \showarticletitle{Qwen technical report}.
\newblock \bibinfo{journal}{\emph{arXiv preprint arXiv:2309.16609}} (\bibinfo{year}{2023}).
\newblock


\bibitem[Choi et~al\mbox{.}(2024)]%
        {choi2024multi}
\bibfield{author}{\bibinfo{person}{Minjin Choi}, \bibinfo{person}{Hye-young Kim}, \bibinfo{person}{Hyunsouk Cho}, {and} \bibinfo{person}{Jongwuk Lee}.} \bibinfo{year}{2024}\natexlab{}.
\newblock \showarticletitle{Multi-intent-aware Session-based Recommendation}. In \bibinfo{booktitle}{\emph{Proceedings of the 47th International ACM SIGIR Conference on Research and Development in Information Retrieval}}. \bibinfo{pages}{2532--2536}.
\newblock


\bibitem[Chowdhary and Chowdhary(2020)]%
        {chowdhary2020natural}
\bibfield{author}{\bibinfo{person}{KR1442 Chowdhary} {and} \bibinfo{person}{KR Chowdhary}.} \bibinfo{year}{2020}\natexlab{}.
\newblock \showarticletitle{Natural language processing}.
\newblock \bibinfo{journal}{\emph{Fundamentals of artificial intelligence}} (\bibinfo{year}{2020}), \bibinfo{pages}{603--649}.
\newblock


\bibitem[Dadoun and Troncy(2020)]%
        {dadoun2020many}
\bibfield{author}{\bibinfo{person}{Amine Dadoun} {and} \bibinfo{person}{Rapha{\"e}l Troncy}.} \bibinfo{year}{2020}\natexlab{}.
\newblock \showarticletitle{Many-to-one recurrent neural network for session-based recommendation}.
\newblock \bibinfo{journal}{\emph{arXiv preprint arXiv:2008.11136}} (\bibinfo{year}{2020}).
\newblock


\bibitem[Devlin(2018)]%
        {devlin2018bert}
\bibfield{author}{\bibinfo{person}{Jacob Devlin}.} \bibinfo{year}{2018}\natexlab{}.
\newblock \showarticletitle{Bert: Pre-training of deep bidirectional transformers for language understanding}.
\newblock \bibinfo{journal}{\emph{arXiv preprint arXiv:1810.04805}} (\bibinfo{year}{2018}).
\newblock


\bibitem[Dias and Fonseca(2013)]%
        {dias2013improving}
\bibfield{author}{\bibinfo{person}{Ricardo Dias} {and} \bibinfo{person}{Manuel~J Fonseca}.} \bibinfo{year}{2013}\natexlab{}.
\newblock \showarticletitle{Improving music recommendation in session-based collaborative filtering by using temporal context}. In \bibinfo{booktitle}{\emph{2013 IEEE 25th international conference on tools with artificial intelligence}}. IEEE, \bibinfo{pages}{783--788}.
\newblock


\bibitem[Diederik(2014)]%
        {diederik2014adam}
\bibfield{author}{\bibinfo{person}{P~Kingma Diederik}.} \bibinfo{year}{2014}\natexlab{}.
\newblock \showarticletitle{Adam: A method for stochastic optimization}.
\newblock \bibinfo{journal}{\emph{(No Title)}} (\bibinfo{year}{2014}).
\newblock


\bibitem[Guo et~al\mbox{.}(2022)]%
        {guo2022learning}
\bibfield{author}{\bibinfo{person}{Jiayan Guo}, \bibinfo{person}{Yaming Yang}, \bibinfo{person}{Xiangchen Song}, \bibinfo{person}{Yuan Zhang}, \bibinfo{person}{Yujing Wang}, \bibinfo{person}{Jing Bai}, {and} \bibinfo{person}{Yan Zhang}.} \bibinfo{year}{2022}\natexlab{}.
\newblock \showarticletitle{Learning multi-granularity consecutive user intent unit for session-based recommendation}. In \bibinfo{booktitle}{\emph{Proceedings of the fifteenth ACM International conference on web search and data mining}}. \bibinfo{pages}{343--352}.
\newblock


\bibitem[Guo et~al\mbox{.}(2024)]%
        {guo2024integrating}
\bibfield{author}{\bibinfo{person}{Naicheng Guo}, \bibinfo{person}{Hongwei Cheng}, \bibinfo{person}{Qianqiao Liang}, \bibinfo{person}{Linxun Chen}, {and} \bibinfo{person}{Bing Han}.} \bibinfo{year}{2024}\natexlab{}.
\newblock \showarticletitle{Integrating Large Language Models with Graphical Session-Based Recommendation}.
\newblock \bibinfo{journal}{\emph{arXiv preprint arXiv:2402.16539}} (\bibinfo{year}{2024}).
\newblock


\bibitem[He et~al\mbox{.}(2016)]%
        {he2016fast}
\bibfield{author}{\bibinfo{person}{Xiangnan He}, \bibinfo{person}{Hanwang Zhang}, \bibinfo{person}{Min-Yen Kan}, {and} \bibinfo{person}{Tat-Seng Chua}.} \bibinfo{year}{2016}\natexlab{}.
\newblock \showarticletitle{Fast matrix factorization for online recommendation with implicit feedback}. In \bibinfo{booktitle}{\emph{Proceedings of the 39th International ACM SIGIR conference on Research and Development in Information Retrieval}}. \bibinfo{pages}{549--558}.
\newblock


\bibitem[Hidasi(2015)]%
        {hidasi2015session}
\bibfield{author}{\bibinfo{person}{B Hidasi}.} \bibinfo{year}{2015}\natexlab{}.
\newblock \showarticletitle{Session-based Recommendations with Recurrent Neural Networks}.
\newblock \bibinfo{journal}{\emph{arXiv preprint arXiv:1511.06939}} (\bibinfo{year}{2015}).
\newblock


\bibitem[Hidasi and Karatzoglou(2018)]%
        {hidasi2018recurrent}
\bibfield{author}{\bibinfo{person}{Bal{\'a}zs Hidasi} {and} \bibinfo{person}{Alexandros Karatzoglou}.} \bibinfo{year}{2018}\natexlab{}.
\newblock \showarticletitle{Recurrent neural networks with top-k gains for session-based recommendations}. In \bibinfo{booktitle}{\emph{Proceedings of the 27th ACM international conference on information and knowledge management}}. \bibinfo{pages}{843--852}.
\newblock


\bibitem[Kullback and Leibler(1951)]%
        {kullback1951information}
\bibfield{author}{\bibinfo{person}{Solomon Kullback} {and} \bibinfo{person}{Richard~A Leibler}.} \bibinfo{year}{1951}\natexlab{}.
\newblock \showarticletitle{On information and sufficiency}.
\newblock \bibinfo{journal}{\emph{The annals of mathematical statistics}} \bibinfo{volume}{22}, \bibinfo{number}{1} (\bibinfo{year}{1951}), \bibinfo{pages}{79--86}.
\newblock


\bibitem[Li et~al\mbox{.}(2017)]%
        {li2017neural}
\bibfield{author}{\bibinfo{person}{Jing Li}, \bibinfo{person}{Pengjie Ren}, \bibinfo{person}{Zhumin Chen}, \bibinfo{person}{Zhaochun Ren}, \bibinfo{person}{Tao Lian}, {and} \bibinfo{person}{Jun Ma}.} \bibinfo{year}{2017}\natexlab{}.
\newblock \showarticletitle{Neural attentive session-based recommendation}. In \bibinfo{booktitle}{\emph{Proceedings of the 2017 ACM on Conference on Information and Knowledge Management}}. \bibinfo{pages}{1419--1428}.
\newblock


\bibitem[Liang et~al\mbox{.}(2016)]%
        {liang2016factorization}
\bibfield{author}{\bibinfo{person}{Dawen Liang}, \bibinfo{person}{Jaan Altosaar}, \bibinfo{person}{Laurent Charlin}, {and} \bibinfo{person}{David~M Blei}.} \bibinfo{year}{2016}\natexlab{}.
\newblock \showarticletitle{Factorization meets the item embedding: Regularizing matrix factorization with item co-occurrence}. In \bibinfo{booktitle}{\emph{Proceedings of the 10th ACM conference on recommender systems}}. \bibinfo{pages}{59--66}.
\newblock


\bibitem[Luo et~al\mbox{.}(2020)]%
        {luo2020collaborative}
\bibfield{author}{\bibinfo{person}{Anjing Luo}, \bibinfo{person}{Pengpeng Zhao}, \bibinfo{person}{Yanchi Liu}, \bibinfo{person}{Fuzhen Zhuang}, \bibinfo{person}{Deqing Wang}, \bibinfo{person}{Jiajie Xu}, \bibinfo{person}{Junhua Fang}, {and} \bibinfo{person}{Victor~S Sheng}.} \bibinfo{year}{2020}\natexlab{}.
\newblock \showarticletitle{Collaborative Self-Attention Network for Session-based Recommendation.}. In \bibinfo{booktitle}{\emph{IJCAI}}. \bibinfo{pages}{2591--2597}.
\newblock


\bibitem[Luo et~al\mbox{.}(2024)]%
        {luo2024integrating}
\bibfield{author}{\bibinfo{person}{Sichun Luo}, \bibinfo{person}{Yuxuan Yao}, \bibinfo{person}{Bowei He}, \bibinfo{person}{Yinya Huang}, \bibinfo{person}{Aojun Zhou}, \bibinfo{person}{Xinyi Zhang}, \bibinfo{person}{Yuanzhang Xiao}, \bibinfo{person}{Mingjie Zhan}, {and} \bibinfo{person}{Linqi Song}.} \bibinfo{year}{2024}\natexlab{}.
\newblock \showarticletitle{Integrating large language models into recommendation via mutual augmentation and adaptive aggregation}.
\newblock \bibinfo{journal}{\emph{arXiv preprint arXiv:2401.13870}} (\bibinfo{year}{2024}).
\newblock


\bibitem[Moreira et~al\mbox{.}(2020)]%
        {moreira2020hybrid}
\bibfield{author}{\bibinfo{person}{Gabriel de Souza~P Moreira}, \bibinfo{person}{Dietmar Jannach}, {and} \bibinfo{person}{Adilson~Marques da Cunha}.} \bibinfo{year}{2020}\natexlab{}.
\newblock \showarticletitle{Hybrid session-based news recommendation using recurrent neural networks}.
\newblock \bibinfo{journal}{\emph{arXiv preprint arXiv:2006.13063}} (\bibinfo{year}{2020}).
\newblock


\bibitem[Oord et~al\mbox{.}(2018)]%
        {oord2018representation}
\bibfield{author}{\bibinfo{person}{Aaron van~den Oord}, \bibinfo{person}{Yazhe Li}, {and} \bibinfo{person}{Oriol Vinyals}.} \bibinfo{year}{2018}\natexlab{}.
\newblock \showarticletitle{Representation learning with contrastive predictive coding}.
\newblock \bibinfo{journal}{\emph{arXiv preprint arXiv:1807.03748}} (\bibinfo{year}{2018}).
\newblock


\bibitem[Ozbay et~al\mbox{.}(2024)]%
        {ozbay2024gnn}
\bibfield{author}{\bibinfo{person}{Begum Ozbay}, \bibinfo{person}{Resul Tugay}, {and} \bibinfo{person}{Sule~Gunduz Oguducu}.} \bibinfo{year}{2024}\natexlab{}.
\newblock \showarticletitle{A GNN Model with Adaptive Weights for Session-Based Recommendation Systems}. In \bibinfo{booktitle}{\emph{Proceedings of the 2024 9th International Conference on Machine Learning Technologies}}. \bibinfo{pages}{258--264}.
\newblock


\bibitem[Park et~al\mbox{.}(2011)]%
        {park2011session}
\bibfield{author}{\bibinfo{person}{Sung~Eun Park}, \bibinfo{person}{Sangkeun Lee}, {and} \bibinfo{person}{Sang-goo Lee}.} \bibinfo{year}{2011}\natexlab{}.
\newblock \showarticletitle{Session-based collaborative filtering for predicting the next song}. In \bibinfo{booktitle}{\emph{2011 First ACIS/JNU International Conference on Computers, Networks, Systems and Industrial Engineering}}. IEEE, \bibinfo{pages}{353--358}.
\newblock


\bibitem[Peng and Zhang(2022)]%
        {peng2022gc}
\bibfield{author}{\bibinfo{person}{Dunlu Peng} {and} \bibinfo{person}{Shuo Zhang}.} \bibinfo{year}{2022}\natexlab{}.
\newblock \showarticletitle{GC--HGNN: A global-context supported hypergraph neural network for enhancing session-based recommendation}.
\newblock \bibinfo{journal}{\emph{Electronic Commerce Research and Applications}}  \bibinfo{volume}{52} (\bibinfo{year}{2022}), \bibinfo{pages}{101129}.
\newblock


\bibitem[Qiao et~al\mbox{.}(2024)]%
        {qiao2024llm4sbr}
\bibfield{author}{\bibinfo{person}{Shutong Qiao}, \bibinfo{person}{Chen Gao}, \bibinfo{person}{Junhao Wen}, \bibinfo{person}{Wei Zhou}, \bibinfo{person}{Qun Luo}, \bibinfo{person}{Peixuan Chen}, {and} \bibinfo{person}{Yong Li}.} \bibinfo{year}{2024}\natexlab{}.
\newblock \showarticletitle{LLM4SBR: A Lightweight and Effective Framework for Integrating Large Language Models in Session-based Recommendation}.
\newblock \bibinfo{journal}{\emph{arXiv preprint arXiv:2402.13840}} (\bibinfo{year}{2024}).
\newblock


\bibitem[Ren et~al\mbox{.}(2024)]%
        {ren2024representation}
\bibfield{author}{\bibinfo{person}{Xubin Ren}, \bibinfo{person}{Wei Wei}, \bibinfo{person}{Lianghao Xia}, \bibinfo{person}{Lixin Su}, \bibinfo{person}{Suqi Cheng}, \bibinfo{person}{Junfeng Wang}, \bibinfo{person}{Dawei Yin}, {and} \bibinfo{person}{Chao Huang}.} \bibinfo{year}{2024}\natexlab{}.
\newblock \showarticletitle{Representation learning with large language models for recommendation}. In \bibinfo{booktitle}{\emph{Proceedings of the ACM on Web Conference 2024}}. \bibinfo{pages}{3464--3475}.
\newblock


\bibitem[Rendle et~al\mbox{.}(2010)]%
        {rendle2010factorizing}
\bibfield{author}{\bibinfo{person}{Steffen Rendle}, \bibinfo{person}{Christoph Freudenthaler}, {and} \bibinfo{person}{Lars Schmidt-Thieme}.} \bibinfo{year}{2010}\natexlab{}.
\newblock \showarticletitle{Factorizing personalized markov chains for next-basket recommendation}. In \bibinfo{booktitle}{\emph{Proceedings of the 19th international conference on World wide web}}. \bibinfo{pages}{811--820}.
\newblock


\bibitem[Schafer et~al\mbox{.}(2007)]%
        {schafer2007collaborative}
\bibfield{author}{\bibinfo{person}{J~Ben Schafer}, \bibinfo{person}{Dan Frankowski}, \bibinfo{person}{Jon Herlocker}, {and} \bibinfo{person}{Shilad Sen}.} \bibinfo{year}{2007}\natexlab{}.
\newblock \showarticletitle{Collaborative filtering recommender systems}.
\newblock In \bibinfo{booktitle}{\emph{The adaptive web: methods and strategies of web personalization}}. \bibinfo{publisher}{Springer}, \bibinfo{pages}{291--324}.
\newblock


\bibitem[Shani et~al\mbox{.}(2005)]%
        {shani2005mdp}
\bibfield{author}{\bibinfo{person}{Guy Shani}, \bibinfo{person}{David Heckerman}, \bibinfo{person}{Ronen~I Brafman}, {and} \bibinfo{person}{Craig Boutilier}.} \bibinfo{year}{2005}\natexlab{}.
\newblock \showarticletitle{An MDP-based recommender system.}
\newblock \bibinfo{journal}{\emph{Journal of machine Learning research}} \bibinfo{volume}{6}, \bibinfo{number}{9} (\bibinfo{year}{2005}).
\newblock


\bibitem[Sun et~al\mbox{.}(2019)]%
        {sun2019self}
\bibfield{author}{\bibinfo{person}{Shiming Sun}, \bibinfo{person}{Yuanhe Tang}, \bibinfo{person}{Zemei Dai}, {and} \bibinfo{person}{Fu Zhou}.} \bibinfo{year}{2019}\natexlab{}.
\newblock \showarticletitle{Self-attention network for session-based recommendation with streaming data input}.
\newblock \bibinfo{journal}{\emph{IEEE Access}}  \bibinfo{volume}{7} (\bibinfo{year}{2019}), \bibinfo{pages}{110499--110509}.
\newblock


\bibitem[Sun et~al\mbox{.}(2024)]%
        {sun2024large}
\bibfield{author}{\bibinfo{person}{Zhu Sun}, \bibinfo{person}{Hongyang Liu}, \bibinfo{person}{Xinghua Qu}, \bibinfo{person}{Kaidong Feng}, \bibinfo{person}{Yan Wang}, {and} \bibinfo{person}{Yew~Soon Ong}.} \bibinfo{year}{2024}\natexlab{}.
\newblock \showarticletitle{Large language models for intent-driven session recommendations}. In \bibinfo{booktitle}{\emph{Proceedings of the 47th International ACM SIGIR Conference on Research and Development in Information Retrieval}}. \bibinfo{pages}{324--334}.
\newblock


\bibitem[Tan et~al\mbox{.}(2016)]%
        {tan2016improved}
\bibfield{author}{\bibinfo{person}{Yong~Kiam Tan}, \bibinfo{person}{Xinxing Xu}, {and} \bibinfo{person}{Yong Liu}.} \bibinfo{year}{2016}\natexlab{}.
\newblock \showarticletitle{Improved recurrent neural networks for session-based recommendations}. In \bibinfo{booktitle}{\emph{Proceedings of the 1st workshop on deep learning for recommender systems}}. \bibinfo{pages}{17--22}.
\newblock


\bibitem[Touvron et~al\mbox{.}(2023)]%
        {touvron2023llama}
\bibfield{author}{\bibinfo{person}{Hugo Touvron}, \bibinfo{person}{Thibaut Lavril}, \bibinfo{person}{Gautier Izacard}, \bibinfo{person}{Xavier Martinet}, \bibinfo{person}{Marie-Anne Lachaux}, \bibinfo{person}{Timoth{\'e}e Lacroix}, \bibinfo{person}{Baptiste Rozi{\`e}re}, \bibinfo{person}{Naman Goyal}, \bibinfo{person}{Eric Hambro}, \bibinfo{person}{Faisal Azhar}, {et~al\mbox{.}}} \bibinfo{year}{2023}\natexlab{}.
\newblock \showarticletitle{Llama: Open and efficient foundation language models}.
\newblock \bibinfo{journal}{\emph{arXiv preprint arXiv:2302.13971}} (\bibinfo{year}{2023}).
\newblock


\bibitem[Wang et~al\mbox{.}(2022)]%
        {wang2022towards}
\bibfield{author}{\bibinfo{person}{Chenyang Wang}, \bibinfo{person}{Yuanqing Yu}, \bibinfo{person}{Weizhi Ma}, \bibinfo{person}{Min Zhang}, \bibinfo{person}{Chong Chen}, \bibinfo{person}{Yiqun Liu}, {and} \bibinfo{person}{Shaoping Ma}.} \bibinfo{year}{2022}\natexlab{}.
\newblock \showarticletitle{Towards representation alignment and uniformity in collaborative filtering}. In \bibinfo{booktitle}{\emph{Proceedings of the 28th ACM SIGKDD conference on knowledge discovery and data mining}}. \bibinfo{pages}{1816--1825}.
\newblock


\bibitem[Wang et~al\mbox{.}(2021)]%
        {wang2021session}
\bibfield{author}{\bibinfo{person}{Jianling Wang}, \bibinfo{person}{Kaize Ding}, \bibinfo{person}{Ziwei Zhu}, {and} \bibinfo{person}{James Caverlee}.} \bibinfo{year}{2021}\natexlab{}.
\newblock \showarticletitle{Session-based recommendation with hypergraph attention networks}. In \bibinfo{booktitle}{\emph{Proceedings of the 2021 SIAM international conference on data mining (SDM)}}. SIAM, \bibinfo{pages}{82--90}.
\newblock


\bibitem[Wang et~al\mbox{.}(2019)]%
        {wang2019collaborative}
\bibfield{author}{\bibinfo{person}{Meirui Wang}, \bibinfo{person}{Pengjie Ren}, \bibinfo{person}{Lei Mei}, \bibinfo{person}{Zhumin Chen}, \bibinfo{person}{Jun Ma}, {and} \bibinfo{person}{Maarten De~Rijke}.} \bibinfo{year}{2019}\natexlab{}.
\newblock \showarticletitle{A collaborative session-based recommendation approach with parallel memory modules}. In \bibinfo{booktitle}{\emph{Proceedings of the 42nd international ACM SIGIR conference on research and development in information retrieval}}. \bibinfo{pages}{345--354}.
\newblock


\bibitem[Wang et~al\mbox{.}(2024b)]%
        {wang2024llm}
\bibfield{author}{\bibinfo{person}{Xinyuan Wang}, \bibinfo{person}{Liang Wu}, \bibinfo{person}{Liangjie Hong}, \bibinfo{person}{Hao Liu}, {and} \bibinfo{person}{Yanjie Fu}.} \bibinfo{year}{2024}\natexlab{b}.
\newblock \showarticletitle{LLM-Enhanced User-Item Interactions: Leveraging Edge Information for Optimized Recommendations}.
\newblock \bibinfo{journal}{\emph{arXiv preprint arXiv:2402.09617}} (\bibinfo{year}{2024}).
\newblock


\bibitem[Wang et~al\mbox{.}(2024a)]%
        {wang2024re2llm}
\bibfield{author}{\bibinfo{person}{Ziyan Wang}, \bibinfo{person}{Yingpeng Du}, \bibinfo{person}{Zhu Sun}, \bibinfo{person}{Haoyan Chua}, \bibinfo{person}{Kaidong Feng}, \bibinfo{person}{Wenya Wang}, {and} \bibinfo{person}{Jie Zhang}.} \bibinfo{year}{2024}\natexlab{a}.
\newblock \showarticletitle{Re2LLM: Reflective Reinforcement Large Language Model for Session-based Recommendation}.
\newblock \bibinfo{journal}{\emph{arXiv preprint arXiv:2403.16427}} (\bibinfo{year}{2024}).
\newblock


\bibitem[Wang et~al\mbox{.}(2020)]%
        {wang2020global}
\bibfield{author}{\bibinfo{person}{Ziyang Wang}, \bibinfo{person}{Wei Wei}, \bibinfo{person}{Gao Cong}, \bibinfo{person}{Xiao-Li Li}, \bibinfo{person}{Xian-Ling Mao}, {and} \bibinfo{person}{Minghui Qiu}.} \bibinfo{year}{2020}\natexlab{}.
\newblock \showarticletitle{Global context enhanced graph neural networks for session-based recommendation}. In \bibinfo{booktitle}{\emph{Proceedings of the 43rd international ACM SIGIR conference on research and development in information retrieval}}. \bibinfo{pages}{169--178}.
\newblock


\bibitem[Wu et~al\mbox{.}(2019)]%
        {wu2019session}
\bibfield{author}{\bibinfo{person}{Shu Wu}, \bibinfo{person}{Yuyuan Tang}, \bibinfo{person}{Yanqiao Zhu}, \bibinfo{person}{Liang Wang}, \bibinfo{person}{Xing Xie}, {and} \bibinfo{person}{Tieniu Tan}.} \bibinfo{year}{2019}\natexlab{}.
\newblock \showarticletitle{Session-based recommendation with graph neural networks}. In \bibinfo{booktitle}{\emph{Proceedings of the AAAI conference on artificial intelligence}}, Vol.~\bibinfo{volume}{33}. \bibinfo{pages}{346--353}.
\newblock


\bibitem[Xia et~al\mbox{.}(2021)]%
        {xia2021self}
\bibfield{author}{\bibinfo{person}{Xin Xia}, \bibinfo{person}{Hongzhi Yin}, \bibinfo{person}{Junliang Yu}, \bibinfo{person}{Qinyong Wang}, \bibinfo{person}{Lizhen Cui}, {and} \bibinfo{person}{Xiangliang Zhang}.} \bibinfo{year}{2021}\natexlab{}.
\newblock \showarticletitle{Self-supervised hypergraph convolutional networks for session-based recommendation}. In \bibinfo{booktitle}{\emph{Proceedings of the AAAI conference on artificial intelligence}}, Vol.~\bibinfo{volume}{35}. \bibinfo{pages}{4503--4511}.
\newblock


\bibitem[Yu et~al\mbox{.}(2020)]%
        {yu2020tagnn}
\bibfield{author}{\bibinfo{person}{Feng Yu}, \bibinfo{person}{Yanqiao Zhu}, \bibinfo{person}{Qiang Liu}, \bibinfo{person}{Shu Wu}, \bibinfo{person}{Liang Wang}, {and} \bibinfo{person}{Tieniu Tan}.} \bibinfo{year}{2020}\natexlab{}.
\newblock \showarticletitle{TAGNN: Target attentive graph neural networks for session-based recommendation}. In \bibinfo{booktitle}{\emph{Proceedings of the 43rd international ACM SIGIR conference on research and development in information retrieval}}. \bibinfo{pages}{1921--1924}.
\newblock


\bibitem[Yuan et~al\mbox{.}(2021)]%
        {yuan2021dual}
\bibfield{author}{\bibinfo{person}{Jiahao Yuan}, \bibinfo{person}{Zihan Song}, \bibinfo{person}{Mingyou Sun}, \bibinfo{person}{Xiaoling Wang}, {and} \bibinfo{person}{Wayne~Xin Zhao}.} \bibinfo{year}{2021}\natexlab{}.
\newblock \showarticletitle{Dual sparse attention network for session-based recommendation}. In \bibinfo{booktitle}{\emph{Proceedings of the AAAI conference on artificial intelligence}}, Vol.~\bibinfo{volume}{35}. \bibinfo{pages}{4635--4643}.
\newblock


\bibitem[Zhang et~al\mbox{.}(2023)]%
        {zhang2023efficiently}
\bibfield{author}{\bibinfo{person}{Peiyan Zhang}, \bibinfo{person}{Jiayan Guo}, \bibinfo{person}{Chaozhuo Li}, \bibinfo{person}{Yueqi Xie}, \bibinfo{person}{Jae~Boum Kim}, \bibinfo{person}{Yan Zhang}, \bibinfo{person}{Xing Xie}, \bibinfo{person}{Haohan Wang}, {and} \bibinfo{person}{Sunghun Kim}.} \bibinfo{year}{2023}\natexlab{}.
\newblock \showarticletitle{Efficiently leveraging multi-level user intent for session-based recommendation via atten-mixer network}. In \bibinfo{booktitle}{\emph{Proceedings of the sixteenth ACM international conference on web search and data mining}}. \bibinfo{pages}{168--176}.
\newblock


\bibitem[Zheng et~al\mbox{.}(2024)]%
        {zheng2024adapting}
\bibfield{author}{\bibinfo{person}{Bowen Zheng}, \bibinfo{person}{Yupeng Hou}, \bibinfo{person}{Hongyu Lu}, \bibinfo{person}{Yu Chen}, \bibinfo{person}{Wayne~Xin Zhao}, \bibinfo{person}{Ming Chen}, {and} \bibinfo{person}{Ji-Rong Wen}.} \bibinfo{year}{2024}\natexlab{}.
\newblock \showarticletitle{Adapting large language models by integrating collaborative semantics for recommendation}. In \bibinfo{booktitle}{\emph{2024 IEEE 40th International Conference on Data Engineering (ICDE)}}. IEEE, \bibinfo{pages}{1435--1448}.
\newblock


\end{thebibliography}










\end{document}